\documentclass[aps,prb,twocolumn,showpacs,floatfix]{revtex4}%
\usepackage{amsmath,amsfonts,amssymb}
\usepackage{multirow}
\usepackage{xspace}%
\usepackage[english]{babel}

\usepackage{natbib}

\usepackage{graphicx,color,fancyhdr,xspace}
\definecolor{darkgreen}{rgb}{0,0.5,0}
\definecolor{darkblue}{rgb}{0,0,0.2}
\definecolor{purple}{rgb}{0.35,0,0.35}
\definecolor{orange}{rgb}{1,0.5,0}

\RequirePackage[
   hyperindex,colorlinks,bookmarksnumbered,
   plainpages=true,pdfstartview=FitH]{hyperref}
\hypersetup{linkcolor=blue,urlcolor=darkblue,citecolor=darkgreen}
\usepackage[pdfstartview=FitH]{hyperref}

  \newcommand{\Sec}[1]{Sec.~\ref{#1}}
  \newcommand{\SEC}[1]{Section~\ref{#1}}

  \newcommand{\EQ}[1]{Equation~(\ref{#1})}
  \newcommand{\Eq}[1]{Eq.~(\ref{#1})}
  
  \newcommand{\Eqs}[2]{Eqs.~(\ref{#1}) and (\ref{#2})} 

  \newcommand{\FIG}[1]{Figure~\ref{#1}}
  \newcommand{\FIGp}[2]{Figure.~\ref{#1}(#2)}

  \newcommand{\Fig}[1]{Fig.~\ref{#1}}
  \newcommand{\Figp}[2]{Fig.~\ref{#1}(#2)}

  \newcommand{\ie}{\emph{i.e.}\xspace}
  \newcommand{\eg}{\emph{e.g.}\xspace}
  \newcommand{\vs}{\emph{vs.}\xspace}
  \newcommand{\wrt}{\emph{w.r.t.}\xspace}
  \newcommand{\cf}{\emph{cf.}\xspace}
  \newcommand{\lhs}{\emph{l.h.s.}\xspace}
  \newcommand{\rhs}{\emph{r.h.s.}\xspace}

  \newcommand{\Hc}{\ensuremath{\mathrm{H.c.}}}

  \DeclareMathOperator*{\trace}{tr}

  \newcommand{\I}{\ensuremath{\mathrm{I}}}
  \newcommand{\F}{\ensuremath{\mathrm{F}}}
  \newcommand{\K}{\ensuremath{\mathrm{K}}}
  \newcommand{\D}{\ensuremath{\mathrm{D}}}
  \newcommand{\X}{\ensuremath{\mathrm{X}}}

  \newcommand{\NK}{\ensuremath{N_{\K}}\xspace}
  \newcommand{\EK}{\ensuremath{E_{\K}}\xspace}

  \newcommand{\Atensor}{$A$-tensor\xspace}
  \newcommand{\Atensors}{$A$-tensors\xspace}
  \newcommand{\ACtensor}{$A^\ast$-tensor\xspace}

  \newcommand{\dmNRG}{\mbox{DM-NRG}\xspace}
  \newcommand{\tdNRG}{\mbox{TD-NRG}\xspace}

  \newcommand{\fdmNRG}{\mbox{\textsf{fdm}\texttt{NRG}}\xspace}
  \newcommand{\fgrNRG}{\mbox{\textsf{fgr}\texttt{NRG}}\xspace}
  \newcommand{\tdmNRG}{\mbox{\textsf{tdm}\texttt{NRG}}\xspace}

  \newcommand{\fdm}{FDM\xspace}
  \newcommand{\FDM}{\ensuremath{\mathrm{FDM}}\xspace}

  \newcommand{\mycite}[1]{\citet{#1} (\citeyear{#1})}

\begin{document}

\title{Tensor networks  and the numerical renormalization group }
\author{Andreas \surname{Weichselbaum}}
\affiliation{Physics Department, Arnold Sommerfeld Center for
Theoretical Physics, and Center for NanoScience,
Ludwig-Maximilians-Universit\"at, 80333 Munich, Germany }

\begin{abstract}
  The full-density-matrix numerical renormalization group has
  evolved as a systematic and transparent setting for the
  calculation of thermodynamical quantities at arbitrary
  temperatures within the NRG framework. It directly evaluates the
  relevant Lehmann representations based on the complete basis
  sets introduced by \mycite{Anders05}. In addition, specific
  attention is given to the possible feedback from low energy
  physics to high energies by the explicit and careful
  construction of the full thermal density matrix, naturally
  generated over a distribution of energy shells. Specific
  examples are given in terms of spectral functions (\fdmNRG),
  time-dependent NRG (\tdmNRG), Fermi-Golden rule calculations
  (\fgrNRG), as well as the calculation of plain thermodynamic
  expectation values. Furthermore, based on the very fact that, by
  its iterative nature, the NRG eigenstates are naturally
  described in terms of matrix product states, the language of
  tensor networks has proven enormously convenient in the
  description of the underlying algorithmic procedures. This paper
  therefore also provides a detailed introduction and discussion
  of the prototypical NRG calculations in terms of their
  corresponding tensor networks.
\end{abstract}

\date{\today}



\maketitle

\section{Introduction}

The numerical renormalization group (NRG) \cite{Wilson75,
Krishna80I, Bulla08} is the method of choice for quantum impurity
models. These consist of an interacting local system coupled to
non-interacting typically fermionic baths, which in their
combination can give rise to strongly correlated quantum-many-body
effects. Through its renormalization group (RG) ansatz, its
collective finite size spectra provide a concise snapshot of the
physics of a given model from large to smaller energies on a
logarithmic scale. A rich set of NRG analysis is based on these
finite size spectra, including statistical quantities that can be
efficiently computed within a single shell approach at an
essentially discrete set of temperatures tied to a certain energy
shell. Dynamical quantities such as spectral functions, however,
necessarily require to combine data from all energy scales. Since
all NRG iterations contribute to a single final curve, traditionally
it had not been clear how to achieve this in a systematic clean way,
specifically so for finite temperatures.

The calculation of spectral properties within the NRG started with
\mycite{Oliveira81,Oliveira85} in the context of X-ray absorption
spectra. This was extended to spectral functions at zero temperature
by \mycite{Sakai89}. Finite temperature together with transport
properties, finally, was introduced by \mycite{Costi92}. An
occasionally crucial feedback from small to large energy scales
finally was taken care of by the explicit incorporation of the
reduced density matrix for the remainder of the Wilson chain
(\dmNRG) by \mycite{Hofstetter00}. While these methods necessarily
combined data from all NRG iterations to cover the full spectral
range, they did so through heuristic patching schemes. Moreover, in
the case of finite temperature, these methods had been formulated in
a single-shell setup that associates a well-chosen characteristic
temperature that corresponds to the energy scale of this shell. For
a more complete listing of references, see \mycite{Bulla08}.

The possible importance of a true multi-shell framework for
out-of-equilibrium situations had already been pointed out by
\mycite{Costi97}. As it turns out, this can be implemented in a
transparent systematic way using the complete basis sets, which
where introduced by \mycite{Anders05} for the feat of real-time
evolution within the NRG (\tdNRG). This milestone development
allowed for the first time to use the quasi-exact method of NRG to
perform real-time evolution to exponentially long time scales. It
emerged together with other approaches to real-time evolution of
quantum many-body systems such as the DMRG. \cite{White04,Daley04}
While more traditional single-shell formulations of the NRG still
exist for the calculation of dynamical quantities using complete
basis sets, \cite{Anders05,Peters06} the latter, however, turned out
significantly more versatile. \cite{Wb07,Wb11_rho, Wb11_aoc,Toth08,
Moca12} In particular, a clean multi-shell formulation can be
obtained using the full-density-matrix (\FDM) approach to spectral
functions \fdmNRG. \cite{Wb07} As it turns out, this essentially
generalizes the \dmNRG \cite{Hofstetter00} to a clean black-box
algorithm, with the additional benefit that it allows to treat
arbitrary finite temperatures on a completely generic footing.
Importantly, the \FDM approach can be easily adapted to related
dynamical calculations, such as the time dependent NRG (\tdmNRG), or
Fermi-Golden-rule calculations (\fgrNRG), \cite{Tureci11, Latta11,
Muender12} as will be described in detail in this paper.

For the \FDM approach, the underlying matrix product state (MPS)
structure of the NRG\cite{Wb07,Wb09} provides an extremely
convenient framework. It allows for an efficient description of the
necessary iterative \emph{contractions} of larger tensor networks,
\ie summation over shared index spaces. \cite{Schollwoeck11}
Moreover, since this quickly can lead in complex mathematical
expressions if spelled out explicitly in detail, it has proven much
more convenient to use a graphical representation for the resulting
tensor networks. \cite{Schollwoeck11} In this paper, this is dubbed
MPS diagrammatics. It compactly describes the relevant procedures
that need to be performed in the actual numerical simulation, and as
such also represents a central part of this paper.

The paper then is organized as follows: the remainder of this
section gives a brief introduction to the NRG, complete basis sets,
its implication for the \FDM approach, and the corresponding MPS
description. As a corollary, this section also discusses the
intrinsic relation of energy scale separation, efficiency of MPS,
and area laws. \SEC{sec:MPSdiagrammatics} gives a brief introduction
to MPS diagrammatics, and its implications for the NRG.
\SEC{sec:applications} provides a detailed description of the \FDM
algorithms \fdmNRG, \tdmNRG, as well as \fgrNRG in terms of their
MPS diagrams. This also includes further additional corollaries,
such as the generic calculation of thermal expectation values.
\SEC{sec:summary} provides summary and outlook. A short appendix,
finally, comments on the treatment of fermionic signs within tensor
networks, considering that NRG typically deals with fermionic
systems.

\subsection{Numerical renormalization group and quantum impurity
systems}

The generic quantum impurity system (QIS) is described by the
Hamiltonian
\begin{equation}
   \hat{H}_\mathrm{QIS} =
   \underset{ \equiv H_0 }{\underbrace{
      \hat{H}_{\mathrm{imp}} + \hat{H}_\mathrm{cpl}(\{\hat{f}_{0\mu}\})
   }}
 + \hat{H}_\mathrm{bath}
\text{,}\label{eq:QIS}
\end{equation}
which consists of a small quantum system (the \emph{quantum
impurity}) that is coupled to a non-interacting macroscopic
reservoir $\hat{H}_\mathrm{bath} = \sum_{k \mu} \varepsilon_{k
\mu}^{\phantom{\dagger}} \hat{c}_{k \mu}^\dagger \hat{c}_{k
\mu}^{\phantom{\dagger}}$, \eg a Fermi sea. Here $\hat{c}_{k
\mu}^\dagger$ creates a particle in the bath at energy
$\varepsilon_{k\mu}$ with flavor $\mu$, such as spin or channel, and
energy index $k$. Typically, $\varepsilon_{k \mu} \equiv
\varepsilon_{k}$. The state of the bath at the location $\vec{r}=0$
of the impurity is given by $\hat{f}_{0\mu} \equiv
\tfrac{1}{\mathcal{N}} \sum_k V_k\hat{c}_{k \mu}$ with proper
normalization $\mathcal{N}^2\equiv \sum_k V_k^2$. The coefficients
$V_k$ are determined by the hybridization coefficients of the
impurity as specified in the Hamiltonian [\eg see \Eq{eq:SIAM:Hcpl}
below]. The coupling $\hat{H}_\mathrm{cpl}(\{ \hat{f}_{0\mu}\})$
then can act arbitrarily within the impurity system, while it
interacts with the baths only through $\hat{f}_{0\mu}^{(\dagger)}$,
\ie its degrees of freedom at the location of the impurity. Overall,
the Hilbert space of the typically interacting \emph{local
Hamiltonian} $\hat{H}_0$ in \Eq{eq:QIS} is considered small enough
so it can be easily treated exactly numerically.

The presence of interaction enforces the treatment of the full
exponentially large Hilbert space. Within the NRG, this consists of
a systematic state-space decimation procedure based on energy scale
separation. (i) The continuum of states in the bath is coarse
grained relative to the Fermi energy using the discretization
parameter $\Lambda>1$, such that with $W$ the half-bandwidth of a
Fermi sea, this defines a set of intervals $\pm W[
\Lambda^{-(m-z+1)/2}, \Lambda^{-(m-z)/2}]$, 
each of which is eventually described by a single fermionic degree
of freedom only. Here $m$ is a positive integer, with the additional
constant $z\in[0,1[$ introducing an arbitrary shift,
\cite{Oliveira92,Zitko09} to be referred to as $z$-shift. (ii) For
each individual flavor $\mu$ then, the coarse grained bath can be
mapped exactly onto a semi-infinite chain, with the first site
described by $\hat{f}_{0\mu}$ and exponentially decaying hopping
amplitudes $t_n$ along the chain. This one-dimensional linear setup
is called the Wilson chain, \cite{Wilson75}
\begin{eqnarray}
   \hat{H}_N &\equiv& \hat{H}_0
    + \sum_\mu \sum_{n=1}^{N} \bigl( t_{n-1}
      \hat{f}_{n-1,\mu}^\dagger \hat{f}_{n,\mu}^{\phantom\dagger}
      + \mathrm{H.c.}
   \bigr)
\text{,}\label{eq:Wilsonchain:N}
\end{eqnarray}
where $\hat{H}_\mathrm{QIS} \simeq \lim_{N\to\infty}\hat{H}_N$. For
larger $n$, it quickly holds \cite{Zitko09,Wb11_rho}
\begin{align}
  \omega_n \equiv \lim_{n\gg 1} t_{n-1} &=
  \tfrac{\Lambda^{z-1}(\Lambda-1)}{\log \Lambda}
  W\Lambda^{-\tfrac{n}{2}}
\text{,}\label{eq:def-escale}
\end{align}
where $\omega_n$ describes the smallest energy scale of a Wilson
chain including all sites up to and including site $n$ (described by
$\hat{f}_{n\mu}$) for arbitrary $\Lambda$ and $z$-shift. In
practice, all energies at iteration $n$ are rescaled by the energy
scale $\omega_n$ and shifted relative to the ground state energy of
that iteration. This is referred to as rescaled energies.

From the point of view of the impurity, the effects of the bath are
fully captured by the hybridization function $\Gamma(\varepsilon)
\equiv \pi\rho(\varepsilon) V^2(\varepsilon)$, which is assumed
spin-independent. For simplicity, a flat hybridization function is
assumed throughout, \ie $\Gamma(\varepsilon)=\Gamma
\vartheta(W-|\varepsilon|)$, with the discretization following the
prescription of \citet{Zitko09}. If not indicated otherwise, all
energies are specified in units of the (half-)bandwidth, which
implies $W:=1$.

\subsubsection{Single impurity Anderson model}

The prototypical quantum impurity model applicable to the NRG is the
single impurity Anderson model (SIAM). \cite{Friedel54,Friedel58,
Anderson61,Anderson67} It consists of a single interacting fermionic
level (d-level), \ie the impurity,
\begin{subequations}\label{eq:SIAM}
\begin{equation}
   \hat{H}_\mathrm{imp} =
      \sum_\sigma \varepsilon_{d\sigma} \hat{n}_{d\sigma}
      + U \hat{n}_{d\uparrow} \hat{n}_{d\downarrow}
\text{.}\label{eq:SIAM:Himp}
\end{equation}
with level-position $\varepsilon_{d\sigma}$ and onsite interaction
$U$. This impurity is coupled through the hybridization
\begin{align}
   \hat{H}_\mathrm{cpl}
  &= \sum_\sigma \bigl(
     \hat{d}_{\sigma}^{\dagger} 
     \underset{ \equiv \sqrt{\tfrac{2\Gamma}{\pi}}\cdot\hat{f}_{0\sigma}
     }{\underbrace{
       \sum_{k} V_{k\sigma} \hat{c}_{k \sigma}^{\phantom{\dagger}}
     }}
   + \Hc \bigr)
\label{eq:SIAM:Hcpl}
\end{align}
\end{subequations}%
to a single spinful non-interacting fermi sea, with $\Gamma$ the
total hybridization strength. Here $\hat{d}^\dagger_{\sigma}$
($\hat{c}^\dagger_{k\sigma}$) creates an electron with spin $\sigma
\in \{\uparrow,\downarrow\}$ at the d-level (in the bath with energy
index $k$), respectively. Moreover, $\hat{n}_{d\sigma} \equiv
\hat{d}^\dagger_{\sigma} \hat{d}_{\sigma}$, and $\hat{n}_{k \sigma}
\equiv \hat{c}_{k \sigma}^\dagger \hat{c}_{k
\sigma}^{\phantom{\dagger}}$. At average occupation with a single
electron, the model has three physical parameter regimes that can be
accessed by tuning temperature: the free orbital regime (FO) at
large energies allows all states at the impurity from empty to
doubly occupied, the local moment regime (LM) at intermediate
energies with a single electron at the impurity and the empty and
double occupied state at high energy only accessible through virtual
transitions, and the low-energy strong coupling (SC) fixed-point or
Kondo regime, where the local moment is fully screened by the
electrons in the bath into a quantum-many-body singlet.

\subsection{Complete basis sets}

Within the NRG, a complete many-body basis \cite{Anders05} can be
constructed from the state space of the iteratively computed NRG
eigenstates $\hat{H}_n \vert s\rangle_n = E_s^n \vert s\rangle_n$.
 With the NRG stopped at some final length $N$ of the
Wilson chain, the NRG eigenstates \wrt to site $n<N$ can be
complemented by the complete state space of the rest of the chain,
$\vert e\rangle_n$, describing sites $n+1,\ldots,N$. The latter
space will be referred to as the \emph{environment}, which due to
energy scale separation will only weakly affect the states $\vert
s\rangle_n$. The combined states,
\begin{equation}
   \vert s e\rangle_n \equiv \vert s\rangle_n \otimes \vert e\rangle_n
\text{,}\label{NRG:def-sen}
\end{equation}
then span the full Wilson chain. Within the validity of \emph{energy
scale separation}, one obtains \cite{Anders05}
\begin{subequations}\label{NRG:basis}
\begin{equation}
   \hat H_N \vert s e\rangle_n \simeq E_s^{n} \vert s e\rangle_n
\text{,} \label{NRG:eigs}
\end{equation}
\ie the NRG eigenstates at iteration $n<N$ are, to a good
approximation, also eigenstates of the full Wilson chain. This holds
for a reasonably large discretization parameter $\Lambda \gtrsim
1.7$. \cite{Wilson75,Bulla08,Saberi08}

\begin{figure}[tb!]
\begin{center}
\includegraphics[width=1\linewidth]{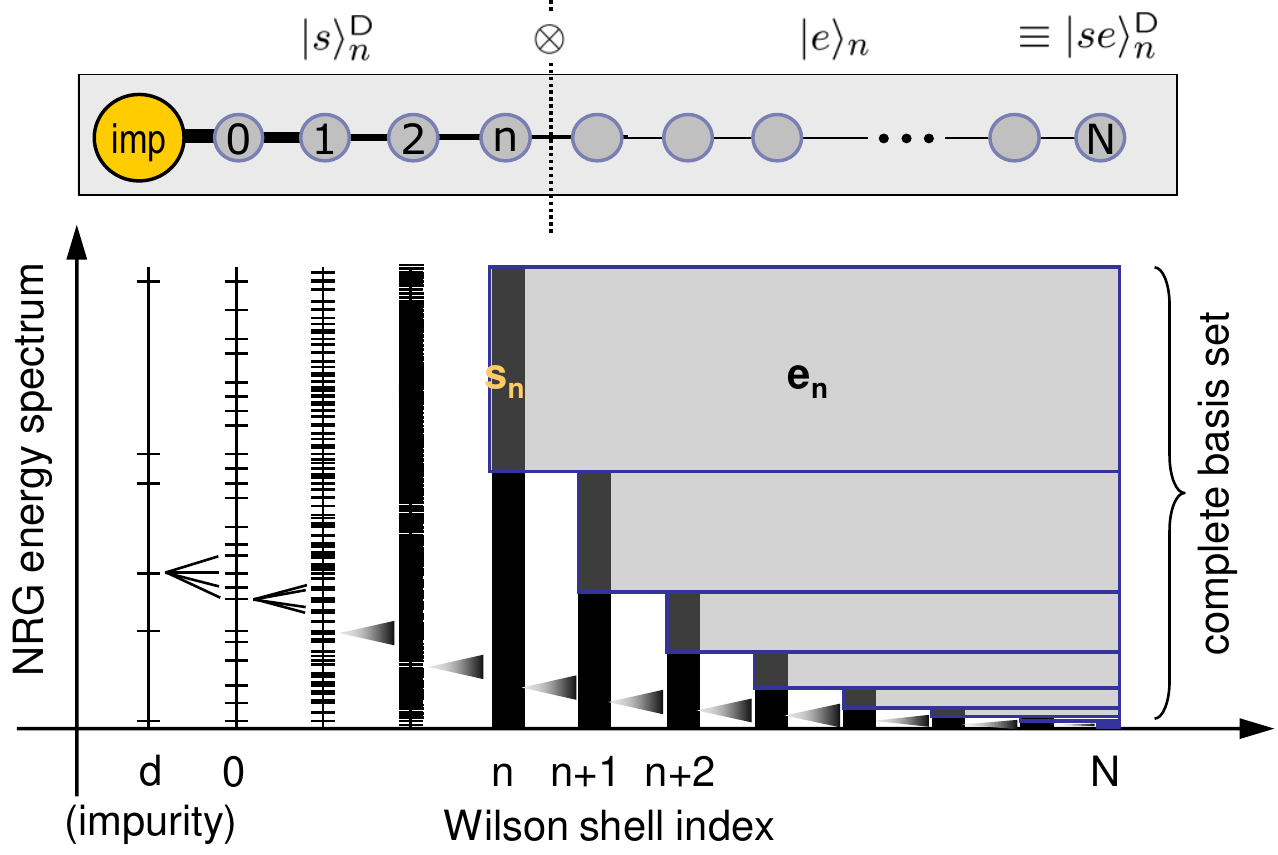}
\end{center}
\caption{
  Iterative construction of complete basis set\cite{Anders05}
  within the NRG by collecting the \emph{discarded} state spaces
  $\vert s\rangle_n^\D$ from all iterations $n\le N$ (black space at
  the left of the gray blocks). For a given iteration $n$, these
  are complimented by the
  environment $\vert e\rangle_n$ for the rest of the system $n'>n$,
  \ie starting from site $n+1$ up
  to the overall chain length $N$ considered (gray blocks). In a
  hand-waving picture, by adding site $n+1$ to the system of sites
  $n'\le n$, this site introduces a new lowest energy scale to the
  system, with the effect that existing levels become split within a
  narrow energy window (indicated by the spread of levels from one
  iteration to the next). The impurity, and also the first few sites
  can be considered exactly with a manageable total dimension of its
  Hilbert space still. Yet as the state space grows exponentially,
  truncation quickly sets in. The \emph{discarded} state spaces then,
  when collected, form a complete basis. At the last iteration, where
  NRG is stopped, by definition, all states are considered discarded.
} \label{fig:nrgbasis}%
\end{figure}

With focus on the iteratively \emph{discarded} state space, this
allows to build a complete many-body \emph{eigen}basis of the full
Hamiltonian, \cite{Anders05}
\begin{equation}
   \mathbf{1}^{( d_0 d^{N}) } =
   \sum_{se,n} \vert se \rangle _{n\ n}^{\D\D}\langle se\vert
\text{,}  \label{NRG:completeness}
\end{equation}
\end{subequations}
where $d_0 d^{N}$ describes the full \emph{many-body} Hilbert space
dimension of the Hamiltonian $H_N$. Here $d$ refers to the state
space dimension of a single Wilson site, while $d_0$ refers to the
state space dimension of the local Hamiltonian $\hat{H}_0$, which in
addition to $\hat{f}_0$ also fully incorporates the impurity [\cf
\Eq{eq:QIS}{}]. It is further assumed, that the local Hamiltonian
$H_0$ is never truncated, \ie truncation sets in for some $n=n_0>0$.
Therefore, by construction, the iterations $n'<n_0$ do not
contribute to \Eq{NRG:completeness}. At the last iteration $n=N$,
all states are considered discarded by definition.\cite{Anders05}
The truncation at intermediate iterations, finally, can be chosen
either \wrt to some threshold number \NK of states to keep, while
nevertheless respecting degenerate subspace, or, preferentially,
\wrt to an energy threshold \EK in rescaled energies [\cf
\Eq{eq:def-escale}{}]. The latter is a dynamical scheme which allows
for a varying number of states depending on the underlying physics.

The completeness of the state space in \Eq{NRG:completeness} can be
easily motivated by realizing that at every NRG truncation step, by
construction, the discarded space (eigenstates at iteration $n$ with
largest energies) is orthogonal to the kept space (eigenstates with
lowest eigenenergies). The subsequent refinement of the kept space
at later iterations will not change the fact, that the discarded
states at iteration $n$ remain orthogonal to the state space
generated at later iterations. This systematic iterative truncation
of Hilbert space while building up a complimentary complete
orthogonal state space is a defining property of the NRG, and as
such depicted schematically in \Fig{fig:nrgbasis}.

\subsection{Identities \label{sec:fdm:ids}}

This section deals with notation and identities related to the
complete basis sets within the NRG. These are essential when
directly dealing with Lehmann representations for the computation of
thermodynamical quantities. While the combination of two basis sets
discussed next simply follows Ref.~\onlinecite{Anders05}, this
section also introduces the required notation. The subsequent
\Sec{sec:fdm:msums} then derives the straightforward generalization
to multiple sums over Wilson shells.

Given the complete basis in \Eq{NRG:basis}, it holds \cite{Anders05}
\begin{equation}
\underset{\equiv \hat{P}_{n}^{\K}}{
\underbrace{\sum_{se} \vert se \rangle _{n~n}^{\K\K} \langle se \vert }}
=
\sum_{n^{\prime }>n}^{N}\underset{\equiv \hat{P}_{n^{\prime }}^{\D}}
{\underbrace{\sum_{se} \vert se \rangle _{n^{\prime }n^{\prime}}^{\D\,\D}
\langle se \vert }}
\text{.}  \label{NRG:sumD2K-0}
\end{equation}%
Here the state space projectors $\hat{P}_{n}^{\X}$ are defined to
project into the kept ($\X=\K$) or discarded ($\X=\D$) space of
Wilson shell $n$. This then allows to rewrite \Eq{NRG:sumD2K-0} more
compactly as
\begin{equation}
   \hat{P}_{n}^{\K}
 = \sum_{n^{\prime }>n}^{(N)} \hat{P}_{n^{\prime }}^{\D}
\text{,} \label{NRG:sumD2K}
\end{equation}%
where the upper limit in the summation, $n'\leq N$, is implied if
not explicitly indicated. With this, two independent sums over
Wilson shells can be reduced into a single sum over shells,
\cite{Anders05}
\begin{eqnarray}
\sum_{n_1,n_2} \hat{P}_{n_1}^{\D} \hat{P}_{n_2}^{\D}
&=&\hspace{-0.14in}
   \sum_{ (n_1 = n_2) \equiv n}\hspace{-0.14in}
   \hat{P}_{n}^{\D} \hat{P}_{n}^{\D}
+  \hspace{-0.14in}\sum_{n_1 > (n_2 \equiv n) }\hspace{-0.14in}
   \hat{P}_{n_1}^{\D} \hat{P}_{n}^{\D}
+  \hspace{-0.14in}\sum_{(n_1 \equiv n) <n_2}\hspace{-0.14in}
   \hat{P}_{n}^{\D}\hat{P}_{n_2}^{\D}
\notag \\
&=&\sum_{n}\left( \hat{P}_{n}^{\D}\hat{P}_{n}^{\D}+\hat{P}_{n}^{\K}\hat{P}%
_{n}^{\D}+\hat{P}_{n}^{\D}\hat{P}_{n}^{\K}\right)
\notag \\
  &\equiv& \underset{
     \equiv {\sum\limits_{n}}^{\prime } }{\underbrace{
     \sum_{n} \sum_{\X\X^\prime}^{\neq \K\K} }}
   \hat{P}_{n}^{\X}\hat{P}_{n}^{\X^{\prime }}
\text{.}  \label{NRG:redSum2to1}
\end{eqnarray}
For simplified notation, the prime in the last single sum over
Wilson shells ($\sum'_n$) indicates that also the kept-sectors are
included in the sum over Wilson shells, yet excluding the all-kept
sector $\X\X' \neq \K\K$, since this sector is refined still in
later iterations. \cite{Anders05,Wb07}

While \Eq{NRG:redSum2to1} holds for the entire Wilson chain, exactly
the same line of arguments can be repeated starting from some
arbitrary but fixed reference shell $n$, leading to
\begin{equation}
  \hat{P}_{n}^{\K}\hat{P}_{n}^{\K}=
  \sum_{n_1, n_2 > n}^{N} \hat{P}_{n_1}^{\D} \hat{P}_{n_2}^{\D}
= \bigr.\sum\limits_{\tilde{n}>n}^{N} \bigr.^{\prime}
  \ \hat{P}_{\tilde{n}}^{\X}\hat{P}_{\tilde{n}}^{\X^{\prime }}
\text{,}\label{NRG:KK2XX}
\end{equation}
where Eq.\ (\ref{NRG:sumD2K}) was used in the first equality.

\subsection{Generalization to multiple sums over shells
\label{sec:fdm:msums}}

Consider the evaluation of some physical correlator that requires
$m>2$ insertions of the identity in \Eq{NRG:completeness} in order
to obtain a simple Lehmann representation. Examples in that respect
are \tdmNRG or (higher-order) correlation functions, as discussed
later in the paper. In all cases, the resulting independent sum over
arbitrarily many identities as in Eq.~(\ref{NRG:completeness}) can
always be rewritten as a \textit{single} sum over Wilson shells. The
latter is desirable since energy differences, such as they occur in
the Lehmann representation for correlation functions, should be
computed within the same shell, where both contributing eigenstates
are described with comparable energy resolution.

\emph{Claim:} Given $m$ full sums as in \Eq{NRG:completeness}, this
can be rewritten in terms of a single sum over a Wilson shell $n$,
such that Eq.~(\ref{NRG:redSum2to1}) generalizes to
\begin{equation}
  \sum_{n_{1},\ldots ,n_{m}}^N \hat{P}_{n_{1}}^{\D_{1}}\ldots
  \hat{P}_{n_{m}}^{\D_{m}}
= \underset{
     \equiv {\sum\limits_{\tilde{n}}}^{\prime}
  }{\underbrace{
     \sum_{\tilde{n}}^N
     \sum_{\X_{1}\cdots \X_{m}}^{\neq \K_{1}\ldots \K_{m}} }}
  \hat{P}_{\tilde{n}}^{\X_{1}}\ldots \hat{P}_{\tilde{n}}^{\X_{m}}
\text{,}\label{NRG:redSumMto1}
\end{equation}
where again the prime in the last single sum over Wilson shells
($\sum'_n$) indicates that \emph{all} states are to be included
within a given iteration $n$, while only excluding the all-kept
sector $\X_1,\ldots,\X_m \neq \K,\ldots,\K$.
Note that via Eq.\ (\ref{NRG:sumD2K}), the \lhs of
Eq.~(\ref{NRG:redSumMto1}) can be rewritten as
\begin{equation*}
  \hat{P}_{n_0-1}^{\K_{1}}\ldots \hat{P}_{n_0-1}^{\K_{m}}
= \sum_{n_{1},\ldots,n_{m}}
  \hat{P}_{n_{1}}^{\D}\ldots \hat{P}_{n_{m}}^{\D}
\end{equation*}%
where $n_{0}>0$ is the first iteration where truncation sets in.
This way, $\hat{P}_{n_0-1}^{\K}$ refers to the full Hilbert space
still. Proving Eq.~(\ref{NRG:redSumMto1}) hence is again equivalent
to proving for general $n$ that
\begin{equation}
   \hat{P}_{n}^{\K_{1}}\ldots \hat{P}_{n}^{\K_{m}}
 = \hspace{-0.16in}\sum_{ n_{1},\ldots,n_{m} >n} \hspace{-0.14in}
   \hat{P}_{n_{1}}^{\D}\ldots \hat{P}_{n_{m}}^{\D}
 = {\sum\limits_{\tilde{n}>n}}\bigr. ^{\prime }
   \ \hat{P}_{\tilde{n}}^{\X_{1}}\ldots \hat{P}_{\tilde{n}}^{\X_{m}}
\text{,} \label{NRG:K2Xgen}
\end{equation}
with the upper limit for each sum over shells, $n_i \leq N$ and
$\tilde{n}\leq N$, implied, as usual. Therefore the sum in the
center term, for example, denotes an independent sum
$\sum_{n_i>n}^N$ for all $n_i$ with $i=1,\ldots,m$.

\emph{Proof:} The case of two sums ($m=2$) was already shown in
Eq.~(\ref{NRG:KK2XX}). Hence one may proceed via induction. Assume,
Eq.~(\ref{NRG:K2Xgen}) holds for $m-1$. Then for the case $m$, one
has in complete analogy to \Eq{NRG:redSum2to1},
\begin{align*}
& \hat{P}_{n}^{\K_{1}}\!\!\ldots \hat{P}_{n}^{\K_{m-1}}\cdot \hat{P}_{n}^{\K_{m}} =\\
& =\Bigl( {\sum\limits_{n'>n}}^{\prime }
   \hat{P}_{n'}^{\X_{1}}\!\!\ldots \hat{P}_{n'}^{\X_{m-1}}\Bigr)
   \Bigl( \sum_{n_{m}>n} \hat{P}_{n_{m}}^{\D_{m}}\Bigr)  \\
&={\sum\limits_{\tilde{n}>n}}^{\!\prime}
    \hat{P}_{\tilde{n}}^{\X_{1}}\!\!\ldots \hat{P}%
   _{\tilde{n}}^{\X_{m-1}}\bigl( \hat{P}_{\tilde{n}}^{\D_{m}}+\hat{P}_{\tilde{n}%
   }^{\K_{m}}\bigr)
 + \hat{P}_{\tilde{n}}^{\K_{1}}\!\!\ldots \hat{P}_{\tilde{n}}^{\K_{m-1}}
   \hat{P}_{\tilde{n}}^{\D_{m}} \\
&\equiv
  {\sum\limits_{\tilde{n}>n}}^{\prime }
  \hat{P}_{\tilde{n}}^{\X_{1}} \!\!\ldots \hat{P}_{\tilde{n}}^{\X_{m}}
\text{,}
\end{align*}%
where from the second to the third line, it was used that,
\begin{align*}
&{\sum\limits_{n'>n}}^{\prime}
  \sum_{n_{m}>n}
= {\sum\limits_{n < (\tilde{n} \equiv  n' = n_{m})}
   }^{\hspace{-0.25in}\prime}\quad
 + { \sum\limits_{n < (\tilde{n}\equiv n') < n_{m}}
    \hspace{-0.25in}}^{\prime }\quad
 + { \sum\limits_{n < (\tilde{n}\equiv n_{m}) < n'}
    \hspace{-0.25in}}^{\prime}\quad
\text{,}
\end{align*}%
and the last term in the third line followed from the inductive
hypothesis. This proves \Eq{NRG:K2Xgen}. $\blacksquare $

Alternatively, the $m$ independent sums over $\{n_1,\ldots n_m\}$ in
Eq.~(\ref{NRG:K2Xgen}) can be rearranged such, that for a specific
iteration $\tilde{n}$, either one of the indices $n_{i}$ may carry
$\tilde{n}$ as minimal value, while all other sums range from
$n_{i^{\prime }}\geq \tilde{n}$. This way, by construction, the
index $n_{i}$ stays within the \textit{discarded} state space, while
all other sums $n_{i'}$ are unconstrained up to $n_{i'}\ge n_i =
\tilde{n}$, thus represent either discarded at iteration $\tilde{n}$
or discarded at any later iteration which corresponds to the kept
space at iteration $\tilde{n}$. From this, \Eq{NRG:K2Xgen} also
immediately follows.

\subsection{Energy scale separation and area laws}

By construction, the iterative procedure of the NRG generates an MPS
representation for its energy eigenbasis. \cite{Wb09,FV_Wb05} This
provides a direct link to the density matrix renormalization group
(DMRG), \cite{White92,Schollwoeck05} and consequently also to its
related concepts of quantum information. \cite{Schollwoeck11} For
example, it can be demonstrated that quite similar to the DMRG, the
NRG truncation \wrt to a fixed energy threshold \EK is also
quasi-variational \wrt to the ground state of the semi-infinite
Wilson chain. \cite{Saberi08, Wb11_rho} Note furthermore that while
DMRG typically targets a single global state, namely the ground
state of the full system, at an intermediate step nevertheless it
also must deal with large effective state spaces describing
disconnected parts of the system. This again is very much similar to
the NRG, which at every iteration needs to deal with many states.

Now the success of variational MPS, \ie DMRG, to ground state
calculations of quasi-one-dimensional systems is firmly rooted in
the so-called area law for the entanglement or block entropy $S_A
\equiv \trace(- \hat{\rho}_A \log \hat{\rho_A})$ with $\hat{\rho}_A
= \trace_B(\hat{\rho})$. \cite{Verstraete08,Wolf08,Schuch08} In
particular, the block entropy $S_A$ represents the entanglement of
some contiguous region $A$ with the rest $B$ of the entire system $A
\cup B$ considered. This allows to explain, why MPS, indeed, is
ideally suited to efficiently capture ground state properties for
quasi-one-dimensional systems.

In constrast to DMRG for real-space lattices, however, NRG
references all energy scales through its iterative diagonalization
scheme. It zooms in towards the low energy scales (``\emph{ground
state properties}'') of the full semi-infinite Wilson chain.
Therefore given a Wilson chain of sufficient length $N$, without
restricting the case, one may consider the fully mixed density
matrix built from the ground state space $\vert 0\rangle_N$ of the
last iteration, for simplicity. This then allows to analyze the
entanglement entropy $S_n$ of the states $\vert s\rangle_n$, \ie the
block of sites $n'<n$, with respect to its environment $\vert
e\rangle_n$.
The interesting consequence in terms of area law is that one expects
the (close to) lowest entanglement entropy $S_n$ for the stable
low-energy fixed point, while one expects $S_n$ to increase for
higher energies, \ie with \emph{de}creasing Wilson shell index $n$.

\begin{figure}[tbp!]
\begin{center}
\includegraphics[width=1\linewidth]{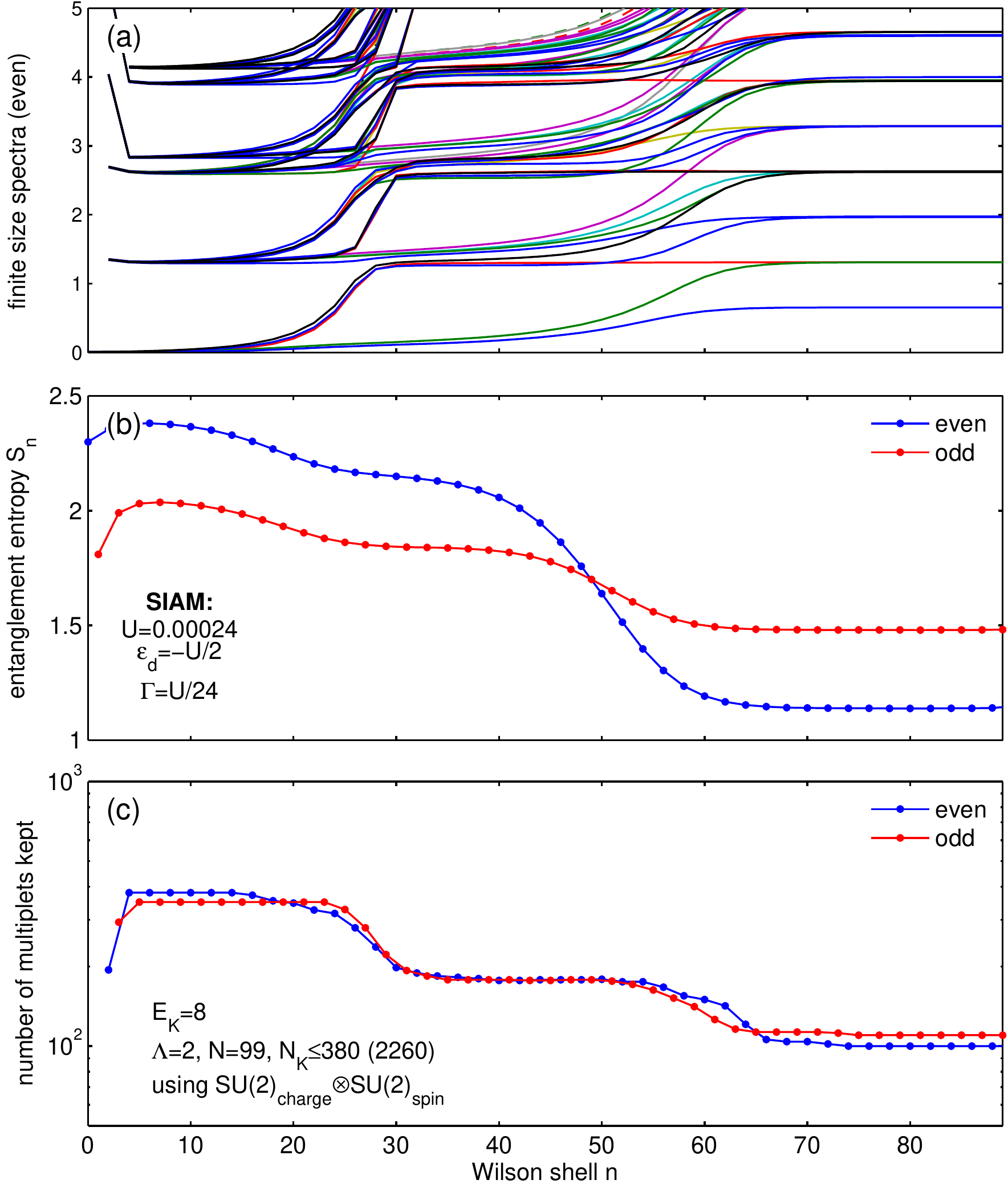}
\end{center}
\caption{
  NRG and area law -- analysis for the symmetric SIAM
  for the parameters as shown in panels (b) and (c)
  [\cf \Eq{eq:SIAM}; all energies in units of bandwidth]. Panel (a)
  shows the standard energy flow diagram of the NRG
  for even iterations where
  the different colors indicate different symmetry sectors. Panel (b)
  shows the entanglement entropy $S_n$ of the Wilson chain up to and
  including site $n<N$ with the rest of the chain, given the overall
  ground state ($N=99$). Due to intrinsic even-odd
  alternations, even and odd iterations $n$ are plotted separately.
  Panel (c) shows the actual number of multiplets kept from one
  iteration to the next, using a dynamical truncation criteria \wrt a
  predefined fixed energy threshold $\EK$ as specified.
  The calculation used
  $\mathrm{SU}(2)_\mathrm{spin}\otimes
   \mathrm{SU}(2)_\mathrm{charge}$ symmetry, hence
  the \emph{actual} number of kept states is by about an order of
  magnitude larger [\eg as indicated with the maximum number of
  multiplets kept, $\NK$ in panel (c): the value in brackets gives the
  corresponding number of states].
}\label{fig:NRG:SEntropy}%
\end{figure}

This is nicely confirmed in a sample calculation for the SIAM, as
demonstrated in \Fig{fig:NRG:SEntropy}. \FIGp{fig:NRG:SEntropy}{a}
shows the standard NRG energy flow diagram (collected finite size
spectra, here for even iterations), which clearly outlines the
physical regimes of free orbital (FO, $n \lesssim 25$), local moment
(LM, $25 \lesssim n \lesssim 60$), and strong coupling (SC, $n
\gtrsim 60$) regime. Here in order to have a sufficiently wide FO
regime, a very small onsite interaction $U$ was chosen relative to
the bandwidth of the Fermi sea.
Panel (b) shows the entanglement entropy $S_n$ between system
($n'\le n$) and environment ($n'>n$). Up to the very beginning or
the very end of the actual chain (the latter is not shown), this
shows a smooth monotonously decaying behavior \vs energy scale. In
particular, consistent with the area law for lowest-energy states,
the entanglement is smallest once the stable low-energy fixed point
is reached. Having chosen a dynamical (quasi-variational)
\cite{Wb11_rho} truncation scheme \wrt to a threshold energy \EK in
rescaled energies [\cf \Eq{eq:def-escale}{}], the qualitative
behavior of the entanglement entropy is also reflected in the number
of states that one has to keep for some fixed overall accuracy, as
shown in \Figp{fig:NRG:SEntropy}{c}. Clearly, up to the very few
first shells prior to truncation, the largest number of states must
be kept at early iterations. While this is a hand-waving argument,
this nevertheless confirms the empirical fact, that the first few
Wilson shells with truncation are usually the most important, \ie
most expensive ones. Therefore for good overall accuracy all the way
down to the low energy sector, one must allow for a sufficiently
large number of states to be kept at early iterations.

The entanglement entropy as introduced above together with the area
law thus is consistent with the energy scale separation along the
Wilson chain in [\cf \FIGp{fig:NRG:SEntropy}{b}]. However, note that
the specific value of the entanglement entropy is not a physical
quantity, in that it depends on the discretization. While the
entanglement entropy clearly converges to a specific value when
including a sufficient number of states, it nevertheless sensitively
depends on $\Lambda$. The smaller $\Lambda$, the larger the
entanglement entropy $S_n$ is going to be, since after all, the
Wilson chain represents a gapless system. The overall qualitative
behavior, however, is expected to remain the same, \ie independent
of $\Lambda$. Similar arguments hold for entanglement spectra and
their corresponding entanglement flow diagram, which provide
significantly more detailed information still about the reduced
density matrices constructed by the bipartition into system and
environment.\cite{Wb11_rho}

\subsection{Full density matrix \label{sec:FDM}}

Given the complete NRG energy eigenbasis $\vert se\rangle_n^\D$, the
full density matrix (\fdm) at arbitrary temperature $T \equiv
1/\beta$ is simply given by \cite{Wb07}
\begin{eqnarray}
  \hat{\rho}^{\FDM}(T) = \sum_{sen}
  \tfrac{e^{-\beta E_s^n}}{Z} \vert se\rangle_{n\,n}^{\D\D}\langle se\vert
\text{,}\label{eq:fdm-rho}
\end{eqnarray}
with $Z(T) \equiv \sum_{ne,s\in\D} e^{-\beta E_s^n}$. By
construction of a thermal density matrix, all energies $E_s^n$ from
all shells $n$ appear on an equal footing relative to a single
global energy reference. Hence any prior iterative rescaling or
shifting of the energies $E_s^n$, which is a common procedure within
the NRG  [\cf \Eq{eq:def-escale}{}], clearly must be undone. From a
numerical point of view, typically the ground state energy at the
last iteration $n=N$ for a given NRG run is taken as energy
reference. In particular, this ensures numerical stability in that
all Boltzmann weights are smaller or equal $1$.

Note that the energies $E_s^n$ are considered \emph{independent} of
the environmental index $e$. As a consequence, this leads to
\emph{exponentially large} degeneracies in energy for the states
$\vert se\rangle_n$. The latter must be properly taken care of
within \fdm, as it contains information from all shells. By already
tracing out the environment for each shell, this leads to
\cite{Wb07}
\begin{eqnarray}
  \hat{\rho}^{\FDM} (T) &=& \sum_{n}
  \underset{ \equiv w_n }{\underbrace{ \tfrac{d^{N-n}Z_n}{Z} \raisebox{-0.18in}{\mbox{}} }}
  \underset{ \equiv \rho_n^{\D}(T) }
  {\underbrace{ \sum_{s} \tfrac{e^{-\beta E_s^n}}{Z_n} \vert s\rangle_{n\,n}^{\D\D}\langle s\vert }}
\text{,} \label{eq:FDM-wn}
\end{eqnarray}
with $d$ the state-space dimension of a single Wilson site, and the
proper normalization by the site-resolved partition function $Z_n(T)
\equiv \sum_{s\in\D_n} e^{-\beta E_s^n}$ of the density matrices
$\rho_n^{\D}(T)$ built from the discarded space of a specific
shell $n$ only. Therefore, $\trace(\rho_n^{\D}(T))=1$, and also
$Z(T)=\sum_n Z_n(T)$. \EQ{eq:FDM-wn} then defines the weights $w_n$,
which themselves represent a normalized distribution, \ie $\sum_n
w_n =1$. Importantly, \Eq{eq:FDM-wn} demonstrates that FDM is
constructed in a well-defined manner from a distribution of density
matrices $\rho_n^{\D}(T)$ from a range of energy shells $n$.

\subsubsection{Weight distribution $w_n$}

The qualitative behavior of the weights $w_n$ can be understood
straightforwardly. With the typical energy scale of shell $n$ given
by
\begin{equation}
   \omega_n = a \Lambda^{-n/2}
\text{,}\label{def:EScale}
\end{equation}
with $a$ some constant of order 1. [\cf \Eq{eq:def-escale}], this
allows to estimate the weights $w_n$ as follows,
\begin{eqnarray*}
   \ln(w_n)
   &\simeq& \ln\bigl(d^{N-n} e^{-\beta\omega_n } / Z\bigr) \notag\\
   &=& (N-n)\ln(d) -\beta\omega_n + \mathrm{const}
\text{.}\label{eq:omega:n}
\end{eqnarray*}
For a given temperature $T$, the shell $n$ with maximum weight is
determined by,
\begin{eqnarray*}
   \tfrac{d}{dn} \ln(w_n) \simeq& -\ln(d)
 + \tfrac{a \beta \ln(\Lambda)}{2}\Lambda^{-n/2} \overset{!}{=} 0
\text{,}
\end{eqnarray*}
with the solution
\begin{eqnarray}
   a\Lambda^{-n^\ast/2} \simeq \tfrac{2\ln(d)}{\beta \ln(\Lambda)}
   \sim T
\text{,}\label{eq:rhoNorm:max}
\end{eqnarray}
since the second term is $1/\beta$ times some constant of order $1.$
This shows that the weight distribution $w_n$ is strongly peaked
around the energy scale of given temperature $T$. With $T\equiv
a\Lambda^{-n_T/2}$ and therefore $n_T \simeq n^\ast$, the
distribution decays super-exponentially fast towards larger energy
scales $n \ll n_T $ (dominated by $e^{-\beta\omega_n}$ \emph{with
exponentially increasing} $\omega_n$ with decreasing $n$). Towards
smaller energy scales $n \gg n_T$, on the other hand, the
distribution $w_n$ decays in a plain exponential fashion (dominated
by $d^{-n}$, since with $\beta\omega_n \ll 1$,
$e^{-\beta\omega_n}\to 1$).

\begin{figure}[tb!]
\begin{center}
\includegraphics[width=1\linewidth]{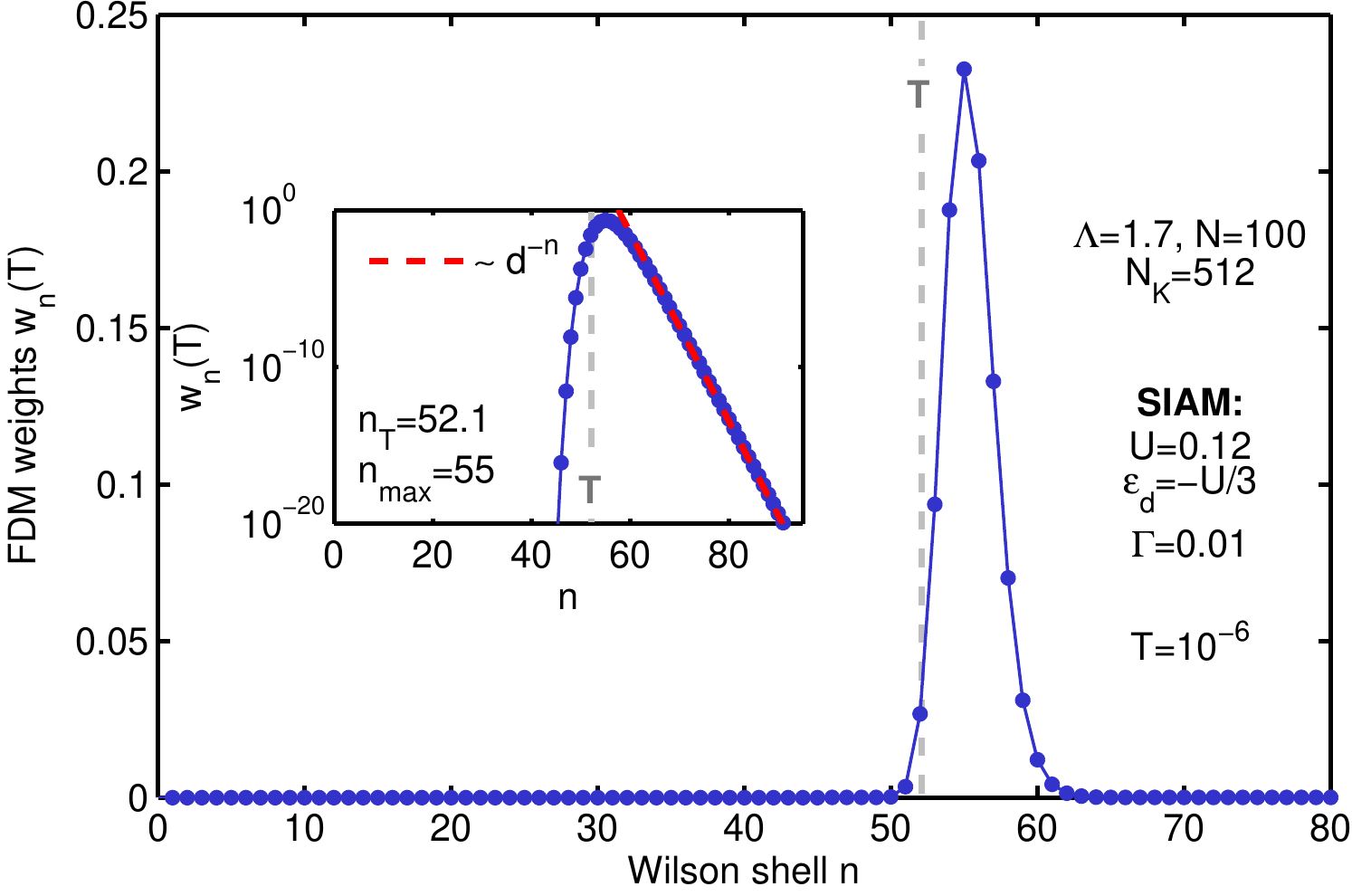}
\end{center}
\caption{
  Typical \fdm weight distribution calculated for the SIAM
  [\cf \Eq{eq:SIAM}] for the parameters as shown in the panel
  and temperature $T=10^{-6}$ (all energies in units of bandwidth).
  The maximum number of states \NK kept at every iteration was
  taken constant. The distribution is strongly peaked around
  the energy shell $n^\ast \gtrsim n_T$, where $n_T$
  (indicated by vertical dashed line) corresponds
  to the energy scale of temperature as defined in the text.
  The inset plots the weights $w_n$ on a logarithmic scale,
  which demonstrates the generic plain exponential decay for small
  energies $n > n_T$, and super-exponentially fast decay towards large
  energies ($n < n_T$).
}\label{fig:rhoNorm}%
\end{figure}

An actual NRG simulation based on the SIAM is shown in
\Fig{fig:rhoNorm}. It clearly supports all of the above qualitative
analysis. It follows for a typical discretization parameter
$\Lambda$ and local dimension $d$, that $n_T$ is slightly smaller
than $n^\ast$, \ie towards larger energies to the left of the
maximum in $w_n$, typically at the left onset of the distribution
$w_n$, as is seen in the main panel in \Fig{fig:rhoNorm} ($n_T$ is
indicated by the vertical dashed line). Within the shell $n^\ast$ of
maximum contribution to the FDM, therefore the actual temperature is
somewhat larger relative to the energy scale of that iteration [note
that this is clearly related to the factor $\bar{\beta}$,
\cite{Krishna80I,Bulla08} introduced by \mycite{Krishna80I}  on
heuristic grounds for the optimal \emph{discrete} temperature
representative for a single energy shell].

An important practical consequence of the exponential decay of the
weights $w_n$ for $n \gg n_T$ is that by taking a long enough Wilson
chain to start with, \fdmNRG \emph{automatically} truncates the
length of the Wilson chain at several iterations past $n_T$.
Therefore the actual length of the Wilson chain $N$ included in a
calculation should be such that the full distribution $w_n$ is
sampled, which implies that $w_n$ has dropped again at least down to
$w_N\lesssim 10^{-3}$.

The weights $w_n$ are fully determined within an NRG calculation,
and clearly depend on the specific physical as well as numerical
parameters. Most obviously, this includes the state space dimension
$d$ of a given Wilson site, and the discretization parameter
$\Lambda$. However, the weights $w_n$ also sensitively depend on the
specific number of states kept from one iteration to the next. For
example, the weights are clearly zero for iterations where no
truncation takes place, which is typically the case for the very
first NRG iterations which include the impurity. However, the
weights also adjust automatically to the specific truncation scheme
adopted, such as the quasi-variational truncation based on an energy
threshold \EK.
In the case of fixed \NK=512 as in \Fig{fig:rhoNorm}, note that if
$d=4$ times the number of states had been kept, \ie $\NK=512 \to
2048$, this essentially would have shifted the entire weight
distribution in \Fig{fig:rhoNorm} by one iteration to the right to
lower energy scales, resulting in an improved spectral resolution
for frequencies $\omega \lesssim T$. \cite{Wb07} For the latter
purpose, however, it is sufficient to use an increased \NK at late
iterations only, where around the energy scale of temperature the
weights $w_n$ contribute mostly.

Furthermore, given a constant number \NK of kept states in
\Fig{fig:rhoNorm}, the weights $w_n$ show a remarkably \emph{smooth}
behavior, irrespective of even or odd iteration $n$. This is
somewhat surprising at first glance, considering that NRG typically
does show pronounced even-odd behavior. For example, for the SIAM
(see also \Fig{fig:NRG:SEntropy}{}), at even iterations an overall
non-degenerate singlet can be formed to represent the ground state.
Having no unpaired spin in the system, this typically lowers the
energy more strongly as compared to odd iterations which do have an
unpaired spin. Therefore while even iterations show a stronger
energy reduction in its low energy states, its ground state space
consists of a single state. In contrast, for odd iterations the
energy reduction by adding the new site is weaker, yet the ground
state space is degenerate, assuming no magnetic field (Kramers
degeneracy). In terms of the corresponding weight distribution for
the full density matrix then, both effects balance each other, such
that distribution of the \fdm weights $w_n$ results in a smooth
function of the iteration $n$, as seen in \Fig{fig:rhoNorm}.

In summary, above analysis shows that the density matrix generated
by \fdm is dominated by several shells around the energy scale of
temperature. The physical information encoded in these shells can
critically affect physical observables at much larger energies. This
construction therefore \emph{shall not} be shortcut in terms of the
density matrix in the kept space at much earlier iterations, \ie by
using $\hat{H}\vert s\rangle_n^\K \simeq E_s^n \vert s\rangle_n^\K$
with the Boltzmann weights thus determined by the energies of the
kept states. This can fail for exactly the reasons already discussed
in detail with the \dmNRG construction by \mycite{Hofstetter00}: the
low-energy physics can have important feedback to larger energy
scales. To be specific, the physics at the low-energy scales on the
order of temperature can play a decisive role on the decay channels
of high-energy excitations. As a result, for example, the low-energy
physics can lead to a significant redistribution of spectral weight
in the local density of states at large energies.

\subsubsection{FDM representation}

The full thermal density matrix $\hat{\rho}_T^{\FDM}$ in
\Eq{eq:FDM-wn} represents a regular operator with an intrinsic
internal sum over Wilson shells. When evaluating thermodynamical
expressions then, as seen through the discussions \Sec{sec:fdm:ids}
and \ref{sec:fdm:msums}, its matrix elements must be calculated both
with respect to discarded \emph{as well as kept states}. While the
former are trivial, the latter require some more attention. All of
this, however, can be written compactly in terms of the projections
in \Eq{NRG:sumD2K-0}.

The reduced density matrix $\hat{\rho}_T^{\FDM}$ is a scalar
operator, from which it follows,
\begin{eqnarray}
  \hat{P}_n^{\X} \hat{\rho}_T^{\FDM} \hat{P}_n^{\X'}
  \equiv \delta_{\X\X'} \hat{R}_n^{\X}
\text{.}\label{eq:spectral:RX}
\end{eqnarray}
This defines the projections $\hat{R}_n^{\X}$ of
$\hat{\rho}_T^{\FDM}$ onto the space $\X \in\{\K,\D\}$ at iteration
$n$, which are not necessarily normalized hence the altered
notation. Like any scalar operator, thus also the projections
$\hat{R}_n^{\X}$ carry a single label $\X$ only. The projection into
the discarded space,
\begin{eqnarray}
   \hat{R}_n^{\D} \equiv \hat{P}_n^{\D} \hat{\rho}_T^{\FDM} \hat{P}_n^{\D}
 = w_n \hat{\rho}_n^{\D}(T)
\text{,}\label{eq:spectral:RD}
\end{eqnarray}
by construction, is a fully diagonal operator as defined in
\Eq{eq:FDM-wn}. In kept space, however, the originally diagonal FDM
acquires \emph{non-diagonal} matrix elements, thus leading to a
non-diagonal scalar operator,
\begin{eqnarray}
   \hat{R}_n^{\K}
 &\equiv& P_n^{\K} \hat{\rho}_T^{\FDM} P_n^{\K} \notag\\
 &=& \sum_{n'>n} w_{n'}
   \underset{ \equiv \hat{\rho}_{n,n'}^{\FDM}(T) }{\underbrace{
       \hat{P}_n^{\K} \hat{\rho}_{n'}^{\D}(T)  \hat{P}_n^{\K}
   }}
\text{,}\label{eq:spectral:RK}
\end{eqnarray}
with the properly normalized reduced density matrices,
\begin{eqnarray}
   \hat{\rho}_{n,n'}^{\FDM}(T) &\equiv&
   \trace_{\{\sigma_{n+1},\ldots,\sigma_{n'}\}}
   \bigl( \hat{\rho}_{n'}^{\D}(T)  \bigr)
\text{.}\label{eq:Rho:nn2}
\end{eqnarray}
These are defined for $n'> n$ and, \wrt to the basis of iteration
$n$, are fully described within its kept space.
Note that in the definition of the $\hat{\rho}_{n'}^{\D}(T)$ in
\Eq{eq:FDM-wn} the \emph{environment} consisting of all sites
$\tilde{n}>n'$ had already been traced out, hence  in
\Eq{eq:Rho:nn2} only the sites $\tilde{n}=n+1,\ldots,n'$ remain to
be considered. By definition, the reduced density matrices
$\rho_{n,n'}^{\FDM}(T)$ are built from the effective basis $\vert
s\rangle_{n'}^\D$ at iteration $n'$, where subsequently the local
state spaces $\sigma_{\tilde{n}}$ of sites $\tilde{n} =
n',n'-1,\ldots, n+1$ are traced out in an iterative fashion.

The projected FDM operators $\hat{R}_n$, like other operators, are
understood as operators in the basis $\vert s\rangle_n$, \ie
$\hat{R}_n^{\X} \equiv \sum_{s\in\X} (R_n^{\X})_{ss'} \vert
s\rangle_{n\,n}\!\langle s'\vert$ (note the hat on the operator),
while the bare matrix elements $(R_n^{\X})_{ss'} \equiv
{}_n\!\langle s\vert \hat{R}_n^{\X} \vert s'\rangle_n$ are
represented by $R_n^{\X}$ (by convention, written without hats).
Overall then, the operator $\hat{R}_n$ can be written in terms of
two contributions, (i) the contribution from iteration $n'=n$ itself
(encoded in discarded space), and (ii) the contributions of all
later iterations $n'>n$ (encoded in kept space at iteration $n$),
\begin{subequations} \label{eq:fdm-Rn}
\begin{eqnarray}
   \hat{R}_n &=&
   \underset{ = \hat{R}_n^\D }{\underbrace{ w_n \hat{\rho}_{n}^{\D}(T)
   \raisebox{-0.19in}{\mbox{}} }}
 + \underset{ = \hat{R}_n^\K }
   {\underbrace{ \sum_{n'> n} w_{n'} \hat{\rho}_{n,n'}^{\FDM}(T) }}
\label{eq:fdm-Rn:1} \\
   &\equiv&
   \sum_{n' \ge n} w_{n'} \hat{\rho}_{n,n'}^{\FDM}(T)
\text{.}\label{eq:fdm-Rn:2}
\end{eqnarray}
\end{subequations}
In the last equation, for simplicity,
the definition of $\hat{\rho}_{n,n'}$ for $n'>n$
in \Eq{eq:Rho:nn2} has been extended to include the case $n'=n$,
where $\hat{\rho}_{n,n} \equiv \hat{\rho}_{n}^{\D}(T)$.

\section{MPS diagrammatics 
\label{sec:MPSdiagrammatics}}

Given the complete basis sets which, to a good approximation, are
also eigenstates of the full Hamiltonian, this allows to evaluate
correlation functions in a text-book like fashion based on their
Lehmann representation. Despite the exponential growth of the
many-body Hilbert space with system size, repeated sums over the
entire Hilbert space nevertheless can be evaluated efficiently, in
practice, due to the one-dimensional structure of the underlying MPS
[the situation is completely analogous to the product, say, of $N$
matrices $A^{(n)}$, $n\in\{1,\ldots,N\}$, of dimension $D$,
$(A^{(1)} A^{(2)} \ldots A^{(N)})_{ij} \equiv \sum_{k_1=1}^D
\sum_{k_2=1}^D \cdots \sum_{k_{N}=1}^D A^{(1)}_{i,k_1}
A^{(2)}_{k_1,k_2} \ldots A^{(N)}_{k_{N-1},j}$. There the sum over
intermediate index spaces $k_1,\ldots,k_{N-1}$ in principle also
grows exponentially with the number of matrices. By performing the
matrix product sequentially, however, this is no problem
whatsoever].

\begin{figure}[tb!]
\begin{center}
\includegraphics[width=\linewidth]{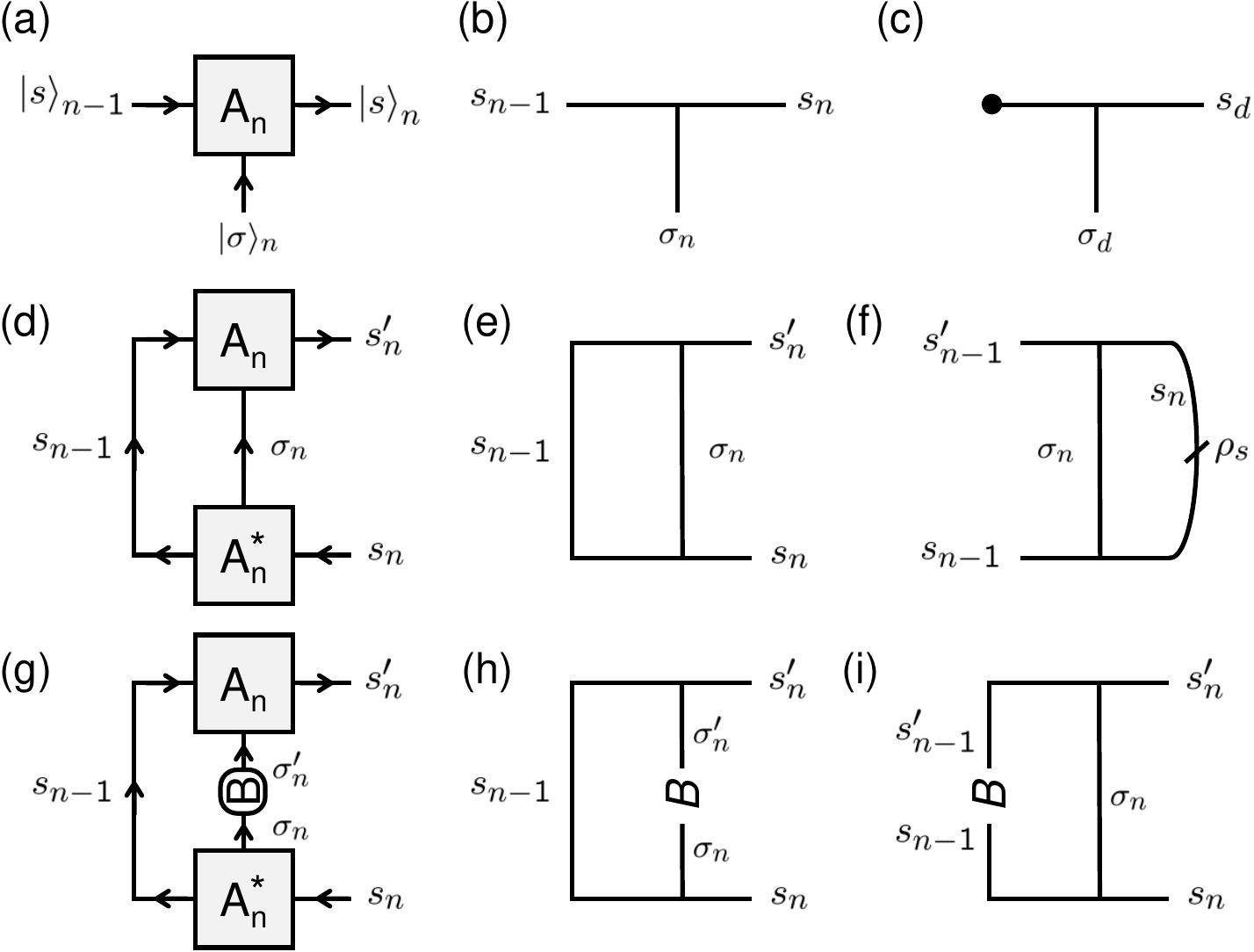}
\end{center}
\caption{
  Basic MPS diagrammatics  -- Panel (a) Iteration step in terms of
  \Atensor. The coefficient space (\Atensor) for given iteration $n$
  is denoted by $A_n$, and incoming and outgoing state spaces are
  indicated by arrows. Panel (b) cleaned up simplified version of
  diagram in panel (a). Panel (c) indicates the first \Atensor, in
  case it has the vacuum state to its left, which is denoted by a
  (terminating) thick dot. Here, trivially $\vert \sigma\rangle_d
  \equiv \vert s\rangle_d$ [with $\vert s\rangle_{0}$ for $n=0$
  generated in the very next iteration with a Wilson chain in mind].
  Panels (d) demonstrates the orthonormality condition of an \Atensor,
  $\sum_{\sigma_n} (A^{[\sigma_n]})^\dagger A^{[\sigma_n]} =
  \mathbf{1}$ [\cf \Eq{eq:mps:ortho}]. Panel (e) again is fully
  equivalent to panel (d). Panel (f) depicts a reduced density matrix.
  Panel (g) represents the evaluation of matrix elements of a local
  operator $\hat{B}$ at site $n$ in the effective state space $s_n$.
  Panel (h) again is a cleaned up simplified version of panel (g).
  Panel (i) is similar to panels (g-h), except that the operator
  $\hat{B}$ was assumed to act at earlier sites on the Wilson chain,
  such that here $B$ already describes the matrix elements in the
  effective basis $s_{n-1}$, and hence contracts from the left.
}\label{fig:mps:basics}%
\begin{center}
\includegraphics[width=\linewidth]{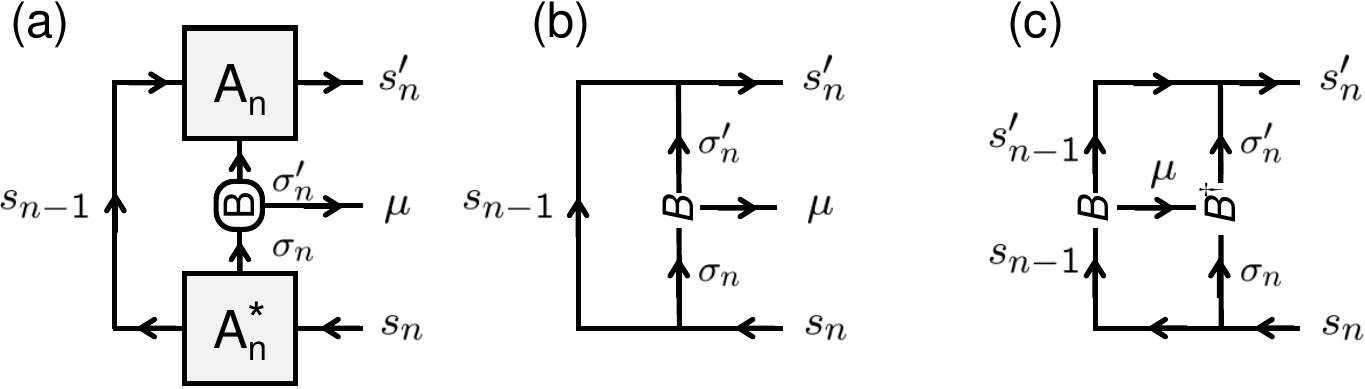}
\end{center}
\caption{
  Basic MPS diagrammatics in the presence of non-abelian symmetries --
  Panel (a) Representation of an irreducible operator $\hat{B}$ that
  acts within the local basis $\sigma_n$ in the effective basis $s_n$.
  Being an irreducible operator, a third open index emerges, both for
  the representation of the local operator $B$ (right incoming index
  to $B_{\sigma_n,\sigma'_n,\mu}  \equiv \langle\sigma_n \vert
  \hat{B}^\mu \vert \sigma'_n\rangle$), as well as for the overall
  contracted effective representation with the open indizes
  $B_{s_n,s'_n,\mu}$, where $\mu$ identifies the spinor component in
  the irreducible operator $\hat{B}$. Panel (b) Simplified version of
  panel (a), but exactly the same otherwise. Panel (c) Contraction
  into a scalar representation of an operator $B$ in the effective
  representation $s_{n-1}$ which acts at some site $n'<n$ with
  operator $B^\dagger$ which acts at site $n$. With $\hat{B} \cdot
  \hat{B}^\dagger \equiv \sum_\mu \hat{B}_\mu \cdot
  \hat{B}^\dagger_\mu$ a scalar operator, the result is a scalar
  operator of rank two in the indices $(s_n,s'_n)$.
  \vspace{-.5in}
}\label{fig:mps:basics2}%
\end{figure}

\subsection{Basics and conventions}

The NRG is based on an iterative scheme: given an (effective)
many-body eigenbasis $\vert s\rangle _{n-1}$ up to and including
site $n-1$ on the Wilson chain, a new site with a $d$-dimensional
state space $\vert \sigma\rangle_n$ is added. Exact diagonalization
of the combined system leads to the new eigenstates
\begin{eqnarray}
   \vert s_{n} \rangle =
   \sum_{s_{n-1},\sigma_{n}} A_{s_{n-1},s_n}^{[\sigma_n]}
   \vert \sigma\rangle_n \vert s\rangle _{n-1}
\text{.}\label{eq:mps:step}
\end{eqnarray}
Here the coefficient space $A_{s_{n-1},s_n}^{[\sigma_n]}$ of the
underlying unitary transformation is already written in standard MPS
notation. \cite{Schollwoeck05,Schollwoeck11} It will be referred to
as \Atensor $A_n$ which, by construction, is of rank $3$.
\EQ{eq:mps:step} is depicted graphically in
\Figp{fig:mps:basics}{a}: two input spaces ($s_{n-1}$ and $\sigma_n$
to the left and at the bottom, respectively), and one output space
$s_{n}$, as indicated by the arrows. Since by convention in this
paper, NRG always proceeds from left to right, \Atensors always have
the same directed structure. Therefore, for simplicity, all arrows
will be skipped later in the paper. Furthermore, the block $A_n$,
which depicts the coefficients of the \Atensor at given iteration,
will be shrunk to a ternary node, resulting in the simplified
elementary building block for MPS diagrams as depicted in panel
\Figp{fig:mps:basics}{b}. Finally, note that the start of the Wilson
chain does not represent any specific specialization. The effective
state space from the previous iteration is simply the vacuum state,
as denoted by the (terminating) thick dot at the left of
\Figp{fig:mps:basics}{c}. The vacuum state represents a perfectly
well-defined and normalized state, such that all subsequent
contractions in the remainder of the panels in \Fig{fig:mps:basics}
apply identically without any specific further modification.

Panel \Figp{fig:mps:basics}{d} depicts the elementary contraction
that represents the orthonormality condition,
\begin{eqnarray}
   \delta_{s_n,s'_n} &=& {}_n\langle s \vert s' \rangle_{n}
   \notag\\
   &=& \sum_{s_{n-1},\sigma_{n}}
         A_{s_{n-1},s'_n}^{[\sigma_n]}
         A_{s_{n-1},s _n}^{[\sigma_n]\ast}
\text{,}\label{eq:mps:ortho}
\end{eqnarray}
again, with panel (e) a cleaned up version, but otherwise exactly
the same as panel (d). By graphical convention, contractions, \ie
summation over shared index or state spaces, are depicted by
lines connecting two tensors. Note that in order to preserve the
directedness of lines in \Figp{fig:mps:basics}{d}, it is important
\wrt bra-states, that all arrows on the \ACtensor belonging to
bra-states are fully reversed. For the remainder of the paper,
however, this is of no further importance.

The contraction in panels \Figp{fig:mps:basics}{d+e} therefore
results in an identity matrix, given that all input spaces of the
\Atensor are contracted. For a mixed contraction, such as one input
and one output state space, on the other hand, as indicated in
\Figp{fig:mps:basics}{f}, this results in a reduced density matrix.
There the sum over the state space $s_n$ is typically weighted by
some normalized, \eg thermal, weight distribution $\rho_s$, as
indicated by the short dash across the line representing $s_n$
together with the corresponding weights $\rho_s$.

Panel \Figp{fig:mps:basics}{g-i} describe matrix representations of
an operator $B$ in the combined effective basis $s_n$ for a local
operator acting within $\sigma_n$ (panels g-h), or for an operator
that acted at some earlier site, such that it already exists in the
matrix representation of the basis $s_{n-1}$. For the latter case,
the contraction in panel (h) typically occurred at some earlier
iteration, with subsequent \emph{iterative} propagation of the
matrix elements as in panel (i) for each later iteration.
Contracting first the the operator $B$ as represented in the state
space of $\sigma'_n$ [$s'_{n-1}$] in \Figp{fig:mps:basics}{h[i]},
respectively, followed by the simultaneous contraction of $(s_{n-1},
\sigma_n)$, the cost of the contraction in panels (g-i) scales like
$\mathcal{O}(D^3)$, where $D$ represents the matrix dimension for
the state spaces $s_{n-1}$ and $s_n$ (here considered to be the
same, for simplicity).

For the NRG it is crucially important to use abelian and non-abelian
symmetries for numerical efficiency. \cite{Wilson75,Wb12_SUN,
Toth08,Moca12, Costi09,Alex11} \FIG{fig:mps:basics2} therefore
presents elementary tensor contractions in the presence of
non-abelian symmetries. \cite{Wb12_SUN} There the basis
transformations in terms of the \Atensors $A_n$ respect the
underlying fusion rules for non-abelian symmetries. Moreover,
elementary operators $\hat{B}$ typically become irreducible operator
sets $\{\hat{B}_\mu\}$ which are described in terms of a spinor with
operator components labeled by the index $\mu$ [see
\Figp{fig:mps:basics2}{a}]. Using Wigner-Eckart theorem, the arrows,
for example, with the operator $\hat{B}$ in
\Figp{fig:mps:basics2}{a} imply the underlying Clebsch-Gordan
coefficient $(\sigma_n| \mu,\sigma'_n)$. \cite{Wb12_SUN} In case of
a scalar operator $\hat{B}$, the spinor reduces to a single
operator, hence $\mu$ reduces to a singleton dimension which can be
stripped. In that case, the third index to the center right of
panels \Figp{fig:mps:basics2}{a-b} can be removed, resulting in the
equivalent diagrams in \Figp{fig:mps:basics}{g-h}. Panel
\Figp{fig:mps:basics2}{c}, finally, shows the contraction of two
irreducible operator sets into a scalar operator $\hat{B} \cdot
\hat{B^\dagger} \equiv \sum_\mu \hat{B}_\mu \cdot
\hat{B}^\dagger_\mu$, again represented in the combined effective
basis $s_n$. Here the operator $\hat{B}$ is considered to have acted
once at some earlier site, whereas its daggered version acts on the
current local site $n$. Note that again the daggered (conjugated)
version has all its arrows reversed where, in addition, in the MPS
diagram the dagger indicates, that the operator $B^\dagger$ as
compared to $B$ has already been also flipped upside down.

The only essential difference when using non-abelian symmetries with
MPS diagrammatics is the emergence of extra indices (lines) \wrt to
irreducible operators (index $\mu$ above). The underlying \Atensors,
of course, need to respect the fusion rules of the symmetries
employed, but on the level of an MPS diagram, this is implied. A
detailed introduction to non-abelian symmetries and its application
to the NRG has been  presented in Ref.~\onlinecite{Wb12_SUN}.
Therefore for the rest of this paper, for simplicity, no further
reference to non-abelian symmetries will be made, with all tensor
networks based on the elementary contractions already presented in
\Fig{fig:mps:basics}.

\section{\fdm Applications \label{sec:applications}}

\subsection{Spectral functions}

Consider the retarded Green's function
\begin{eqnarray}
   G_{BC}^R(t) \equiv -i \vartheta(t)
   \underset{ \equiv G_{BC}(t) }{\underbrace{
   \langle \hat{B}(t) \hat{C}^\dagger\rangle_T }}
\label{eq:G(t)_AB}
\end{eqnarray}
which may regarded as the first term in the standard fermionic
Green's function $G^R(t) = -i \vartheta(t) \langle \{\hat{B}(t)
\hat{C}^\dagger \} \rangle_T$. Here $\hat{B}(t) \equiv e^{i\hat{H}t}
\hat{B} e^{-i\hat{H}t}$, where, as usual, the Hamiltonian $\hat{H}$
of the system is considered time-independent. In \Eq{eq:G(t)_AB}, an
operator $\hat{C}^\dagger$ acts at time $t=0$ on a system in thermal
equilibrium at temperature $T$, described by the thermal density
matrix $\hat{\rho}(T) = e^{-\beta\hat{H}}/Z(T)$ with $\langle \cdot
\rangle_T \equiv \trace\bigl(\hat{\rho}(T) \,\cdot\, \bigr)$. The
system then evolves to some time $t>0$, where a possibly different
operator $\hat{B}$ is applied. The overlap with the original time
evolved wave function then defines the retarded correlation function
of the two events. Fourier-transformed into frequency space, $
G^R(\omega)\equiv \int \tfrac{dt}{2\pi} e^{i\omega t}G^R(t)$, its
spectral function is defined by
\begin{eqnarray}
   A_{BC}(\omega) &\equiv& -\tfrac{1}{\pi}
   \mathrm{Im} G^R_{BC}(\omega)
 = \int \tfrac{dt}{2\pi} e^{i\omega t}G^R_{BC}(t) \notag\\
&=&\int \tfrac{dt}{2\pi} e^{i\omega t}
   \trace \Bigl(
       \hat{\rho}(T) e^{i\hat{H}t} \hat{B} e^{-i\hat{H}t}
       \hat{C}^\dagger
   \Bigr)
\text{.}\label{eq:spectral-def}
\end{eqnarray}
When evaluated in the full many-body eigenbasis, in principle, this
requires the insertion of two identities, (i) to evaluate the trace,
and (ii) in between the operators $\hat{B}$ and $\hat{C}^\dagger$ to
deal with the exponentiated Hamiltonian. For simplified, with the
eigenbasis sets $\mathbf{1} = \sum_a \vert a\rangle\langle a \vert =
\sum_b \vert b\rangle\langle b \vert$, the spectral function
becomes,
\begin{eqnarray}
   A_{BC}(\omega) &=& \sum_{ab}
   \int \tfrac{dt}{2\pi} e^{i(\omega - E_{ab}) t}  \rho_a 
   \langle a\vert \hat{B} \vert b\rangle
   \langle b\vert \hat{C}^\dagger \vert a\rangle \notag \\
&\equiv& \sum_{ab}
   \rho_a B_{ab} C_{ab}^\ast
   \cdot\delta\left(\omega - E_{ab} \right)
\text{,}\label{eq:lehmann-spectral}
\end{eqnarray}
with $E_{ab} \equiv E_{b} - E_{a}$ and $\rho_a \equiv
\tfrac{1}{Z}e^{-\beta E_a}$. By convention, as usual, operators
carry hats, while matrix representations in a given basis have no
hats ($\hat{B}$ \vs $B_{ab}$). \EQ{eq:lehmann-spectral} is referred
to as the Lehmann representation of the correlation function in
\Eq{eq:G(t)_AB}.
In the case of equal operators, $\hat{B} = \hat{C}$, the spectral
function is a strictly positive function, \ie a spectral density. In
either case, the integrated spectral function results in the plain
thermodynamic expectation values,
\begin{eqnarray}
   \int d\omega A_{BC}(\omega)
 = \sum_{ab} \rho_a\, B_{ab} C_{ab}^\ast
 = \bigl\langle \hat{B} \hat{C}^\dagger
   \bigr\rangle_T
\text{.}\label{eq:fdm-spec-xval}
\end{eqnarray}

\begin{figure}[tb!]
\begin{center}
\includegraphics[width=\linewidth]{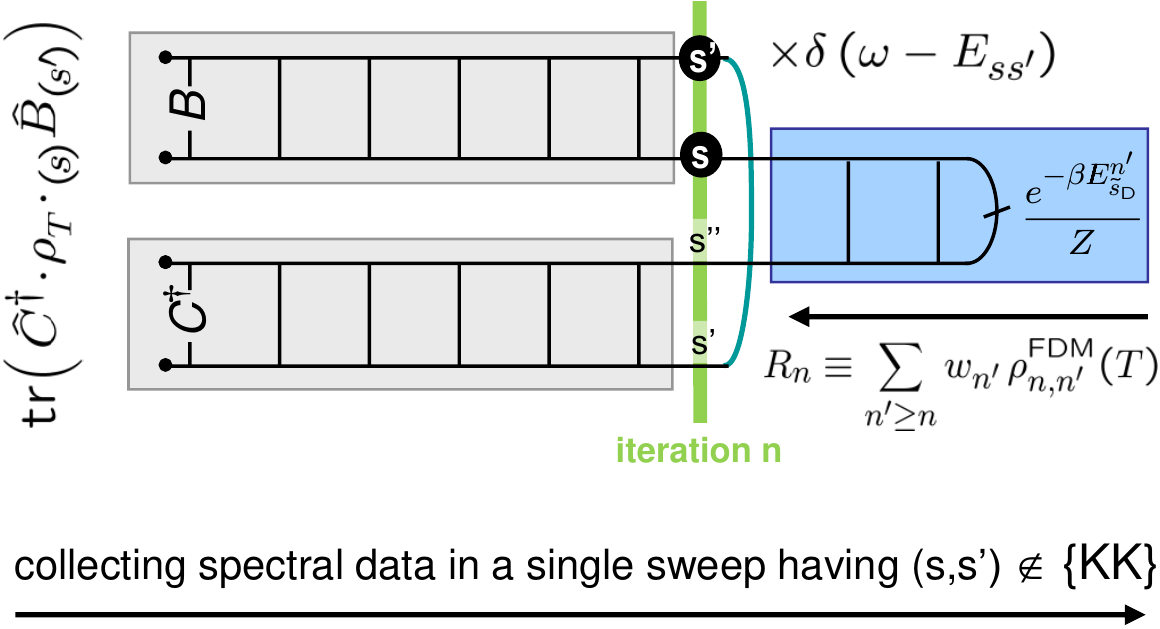}
\end{center}
\caption{ (Color online)
  MPS diagram for calculating spectral functions using \fdmNRG based
  on the Lehmann representation in \Eq{eq:fdm-spec}. In general,
  spectral functions are preceded by an NRG \emph{forward sweep},
  which generates the NRG eigenbasis decomposition (horizontal lines;
  \cf \Fig{fig:mps:basics}). Correlation functions then require the
  evaluation of the matrix elements $\trace\bigl( \hat{C} \rho_{_T}
  \cdot {}_{(s)} \hat{B}_{(s')}\bigr)$, as indicated at the left of
  the figure. The energies of the indices (states) $s$ and $s'$ are
  ``probed'' such that their difference determines the energy $\omega
  = E_{ss'}^n \equiv E_{s'}^n - E_{s}^n$ of an individual contribution
  to the spectral function, as indicated by the $\times \delta(\omega
  - E_{ss'})$ next to the indices $s$ and $s'$ in the upper right of
  the figure. The sum $\sum_{n'>n}$ in the discarded state space of
  $\hat{\rho}^{\FDM}(T)$, indicated to the lower right, results in the
  object $R_n$ [\cf \Eq{eq:fdm-Rn:2}{}]. The individual contributions
  $\rho_{n,n'}^{\FDM}(T)$ are generated by the Boltzmann weights in
  the discarded space at iteration $n'$, as indicated to the right.
  The contribution at $n'=n$, \ie $R_n^\D$, can simply be determined
  when needed. On the other hand, the cumulative contributions $n'>n$
  are obtained in a simple prior \emph{backward sweep}, starting from
  the last Wilson shell $N$ included, as indicated by the small arrow
  pointing to the left. Having $n'>n$, this calculation always maps to
  the kept space, thus resulting in $R_n^\K$. Finally, the spectral
  data is collected in a single forward sweep, as indicated at the
  bottom of the figure.
}\label{fig:fdmnrg}%
\end{figure}

Now using the complete NRG eigenbasis, $\vert a\rangle \to \vert
se\rangle_n$ and $\vert b\rangle \to \vert s'e'\rangle_{n'}$, one
may have been tempted of directly reducing the double sum in
\Eq{eq:lehmann-spectral} to a single sum over Wilson shells using
\Eq{NRG:redSum2to1}. This implies that the thermal weight would be
constructed as $\rho_a(T) \sim e^{-\beta E_a} \to e^{-\beta
E_s^{n,\X}}$ from both, the discarded ($\X=\K$) as well as the kept
($\X=\K$) space at iteration $n$. This, however, ignores a possible
feedback from small to large energy scales which has been shown to
be crucial in the NRG context. \cite{Hofstetter00}

The solution is to take the \fdm as it stands in \Eq{eq:fdm-rho}.
This, however, introduces yet another independent sum $c$ over
Wilson shells, in addition to $a$ and $b$ in
\Eq{eq:lehmann-spectral} above,
\begin{eqnarray}
   A_{BC}(\omega) &=& \sum_{abc}
   \rho_{ca} B_{ab} C_{ac}^\ast
   \cdot\delta\left(\omega - E_{ab} \right)
\text{.}\label{eq:spectral:FDM}
\end{eqnarray}
The triple-sum over $\{a,b,c\}$  can be treated as in
\Eq{NRG:redSumMto1}. With $\{a,b,c\} \to \{s,s',s_\rho\}_n \in
\{\X\X'\X_\rho \neq \K\K\K \}$, nevertheless, $\X=\X_\rho$ are
locked to each other since $\rho$ itself represents a scalar
operator, \ie does not mix kept with discarded states. Therefore
only the contributions $\X\X' \neq \K\K$ as known from a double sum
remain. With
\begin{eqnarray*}
  &&\trace\bigl(
        \hat{\rho}_T^{\FDM}
        \hat{B}(t) 
        \cdot \hat{C}^\dagger
    \bigr) \notag\\
  && = \sum_n \sum_{\X\X'\X_\rho}\trace\bigl(
        \underset{ = \delta_{\X\X_\rho} \hat{R}_n^{\X}}{\underbrace{
          \hat{P}_n^{\X_\rho} \cdot
          \hat{\rho}_T^{\FDM} \cdot \hat{P}_n^{\X}
        }}
        \ \hat{B}(t) 
        \hat{P}_n^{\X'}  \cdot \hat{C}^\dagger
    \bigr)
\text{,}
\end{eqnarray*}
it follows for spectral functions (\fdmNRG),
\begin{eqnarray}
   A_{BC}(\omega) &=& {\sum_{n,ss'}}'
        \bigl[ C^\dagger_n \, R_n \bigr] _{s's}
        (B_n)_{ss'}
      \,\delta\left(\omega - E_{ss'}^n \right)
\text{,}\label{eq:fdm-spec}
\end{eqnarray}
where the prime with the sum again indicates that only the
combinations of states $ss'\in\X\X' \neq \K\K$ are to be considered
at iteration $n$. To be specific, given the scalar nature of the
projections $\hat{R}_n$, the first term implies the matrix product
$( C^\dagger_n R_n)^{\X'\X} \equiv (C^\dagger_n)^{\X'\X} R_n^{\X}$.

The MPS diagram of the underlying tensor structure is shown in
\Fig{fig:fdmnrg}. Every \emph{leg} of the ``ladders'' in
\Fig{fig:fdmnrg} corresponds to an NRG eigenstate (MPS) $\vert
s\rangle_{n}$ for some intermediate iteration $n$. The blocks for
the MPS coefficient spaces ($A$-tensors) are no longer drawn, for
simplicity [\cf \Fig{fig:mps:basics}]. The outer sum over the states
$s'$ in \Eq{eq:fdm-spec} corresponds to the overall trace. Hence the
upper- and lower-most leg in \Fig{fig:fdmnrg} at iteration $n$ carry
the same state label $s'$, as they are connected by a line
(contraction). Furthermore, the inserted identity in the index $s$
initially also would have been identified with two legs [similar to
what is seen in \Fig{fig:tdmnrg} later]. At iteration $n$, however,
the state space $s$ directly hits the \fdm, leading to the overlap
matrix ${}_n^{\X}\langle s \vert \tilde{s}\rangle_n^{\tilde{\X}} =
\delta_{s\tilde{s}} \delta_{\X\tilde{\X}}$ [hence this eliminates
the second block from the top in \Fig{fig:tdmnrg}]. As a result,
only the single index $s$ from the second complete sum remains in
\Fig{fig:fdmnrg}. The same argument applies for the index $s''$.

The two legs in the center of \Fig{fig:fdmnrg}, finally, stem from
the insertion of the \fdm which can extend to all iterations $n'\ge
n$. Note that the case $n'<n$ does not appear, since there the
discarded state space used for the construction of the \fdm is
orthogonal to the state space $s$ at iteration $n$. The trace over
the environment at iteration $n$ leads to the reduced (partial)
density matrices $R_n$. Here the environmental states $\vert e
\rangle_{n'} $ for the density matrices $\rho_{n'}^{\D}(T)$ for
$n'\ge n$ had already all been traced out, as pointed out with
\Eq{eq:FDM-wn}. The \fdm thus reduces at iteration $n$ to the scalar
operator $R_n^{(\X)}$ as introduced in \Eq{eq:fdm-Rn}.

In summary, by insisting on using the \fdm in \Eq{eq:fdm-rho} this
only leads to the minor complication that $R_n^\K$ needs to be
constructed and included in the calculation. The construction of
$R_n^\K$, on the other hand, can be done in a simple prior backward
sweep, which allows to generate $R_n^\K$ iteratively and thus
efficiently. All of the $R_n^\K$ need to be stored for the later
calculation of the correlation function. Living in kept space,
however, the computational overhead is negligible. The actual
spectral data, finally, is collected in a single forward sweep, as
indicated in \Fig{fig:fdmnrg}.

\subsubsection{Exactly conserved sum rules}

By construction, \fdm allows to exactly obey sum-rules for spectral
functions as a direct consequence of \Eq{eq:fdm-spec-xval} and
fundamental quantum mechanical commutator relations. For example,
after completing the Green's function in \Eq{eq:G(t)_AB} to a proper
many-body correlation function for fermions, $G_d(t) \equiv -i
\vartheta(t) \langle \{\hat{d}(t), \hat{d}^\dagger\}\rangle_T$, with
$\hat{d}^\dagger$ creating an electron in level $d$ at the impurity
and $\{\cdot,\cdot\}$ the anticommutator, the integrated spectral
function results in
\begin{eqnarray}
   \int d\omega A(\omega)
 = \bigl\langle \{ \hat{d}, \hat{d}^\dagger\} \bigr\rangle_T = 1
\text{,}\label{eq:sum-rule}
\end{eqnarray}
due to the fundamental fermionic anticommutator relation, $ \{
\hat{d}, \hat{d}^\dagger\} = 1$. In practice, \Eq{eq:sum-rule} is
obeyed exactly within numerical double precision noise ($10^{-16}$),
which underlines the fact that the full exponentially large
quantum-many body state space can be dealt with in practice, indeed.
Note, however, that \Eq{eq:sum-rule} holds by construction, and
therefore it is not measure for convergence of an NRG calculation.
The latter must be checked independently. \cite{Wb11_rho}

\subsubsection{Implications for complex Hamiltonians}

The Hamiltonians analyzed by NRG are usually time-reversal
invariant, and therefore can be computed using non-complex numbers.
In case the Hamiltonian is not time-reversal invariant, \ie the
calculation becomes intrinsically complex, the \Atensors on the
lower leg of the ladders for the operators $\hat{B}$ and
$\hat{C}^\dagger$ in \Fig{fig:fdmnrg} must be complex conjugated
(see also \Fig{fig:mps:basics}). Consequently, this implies for the
\fdm projections $R_n$, that in \Fig{fig:fdmnrg} its constituting
\Atensors in the \emph{upper} leg need to be complex conjugated.

\subsection{Thermal expectation values}

Arbitrary thermodynamic expectation values can be calculated within
the \fdmNRG framework, in priniciple, through \Eq{eq:fdm-spec-xval}.
Given the spectral data on the \lhs of \Eq{eq:fdm-spec-xval}, for
example, this can be integrated to obtain the thermodynamic
expectation value on the \rhs of \Eq{eq:fdm-spec-xval}. In practice,
this corresponds to a simple sum of the non-broadened discrete
spectral data as obtained from \fdmNRG. Using the plain discrete
data has the advantage that it does not depend on any further
details of broadening procedures which typically would introduce
somewhat larger error bars otherwise.

For dynamical properties within the NRG, usually only \emph{local
operators} are of interest. That is, for example, the operators
$\hat{B}$ or $\hat{C}$ in \Eq{eq:fdm-spec-xval} act within the
\emph{local} Hamiltonian $\hat{H}_0$ [Wilson shell $n=0$; \cf
\Eq{eq:QIS}], or within the very first Wilson sites $n < n_0$, where
$n_0$ stands for the first Wilson shell where truncation sets in.
For this early part of the Wilson chain, the weights $w_{n}$ are
identically zero. Consequently, the reduced thermal density matrix
is fully described for iteration $n<n_0$ for arbitrary temperatures
$T$ by $R_n^{\K}(T)$ in kept space. For a given temperature, the
aforementioned simple backward sweep to calculate $R_n^{\K}$ then
already provides all necessary information for the simple evaluation
of the thermal expectation value of any local operator $\hat{C}$
[\eg $\hat{C} := \hat{B} \hat{C}^\dagger$ in \Eq{eq:fdm-spec-xval}],
\begin{eqnarray}
   \bigl\langle \hat{C} \bigr\rangle_T
&=& \trace\bigl[
   R_n^{(\K)}(T) C_{n}^{(\K\K)} \bigr], \qquad (n<n_0)
\label{eq:fdm-xvals}
\end{eqnarray}
with $C_{n}^{(\K\K)}$ the matrix elements of the operator $\hat{C}$
in the kept space of iteration $n$. With no truncation yet at
iteration $n$, the kept space is the only space available, \ie
represents the full state space up to iteration $n$ [hence the
brackets around the $\K$'s]. For strictly local operators acting
within the state space of $H_0$, one has $\langle \hat{C} \rangle_T
= \trace \bigl[ R_0(T) C_{0} \bigr]$. The clear advantage of
\Eq{eq:fdm-xvals} is, that once $R_n^{(\K)}(T)$ has been obtained
for given temperature, \emph{any} local expectation value can be
computed in a simple manner \emph{without} the need to explicitly
calculate the matrix elements of the operator $\hat{C}$ throughout
the entire Wilson chain.

In \Eq{eq:fdm-xvals}, it was assumed that the operator $\hat{C}$
acts on sites $n \leq n_0$ only. This can be relaxed significantly,
however, assuming that temperature is typically much smaller than
the bandwidth of the system. In that case, the weight distribution
$w_n$ already also has absolutely negligible contribution at earlier
iterations $n'\ll n_T$ which clearly stretches beyond $n_0$ (see
\Fig{fig:rhoNorm} and discussion). Hence \Eq{eq:fdm-xvals} can be
relaxed to all iterations $n$ for which $\sum_{n'<n} w_{n'} \ll 1$.

In the case that the operator $\hat{C}$ is not a local operator at
all, but nevertheless acts locally on some specific Wilson site $n$,
then using \Eqs{eq:FDM-wn}{eq:fdm-Rn} it follows for the general
case,
\begin{eqnarray}
   \bigl\langle \hat{C} \bigr\rangle_T
   &=& \trace\bigl[ R_n^{\K}(T) C_{n}^{\K\K} \bigr]
     + \trace\bigl[ R_n^{\D}(T) C_{n}^{\D\D} \bigr]
  \notag \\
    &+& c \sum_{n'<n} w_{n'}
\text{,}\label{eq:FDM:xvals-2}
\end{eqnarray}
which corresponds to the partitioning of \Eq{eq:FDM-wn} given by
$\sum_{n'} = \sum_{n'>n} +  \sum_{n'=n} + \sum_{n'<n}$,
respectively. The last term in \Eq{eq:FDM:xvals-2} derives from the
discarded state spaces for Wilson shells $n'<n$ at (much) larger
energy scales. Therefore the fully mixed thermal average applies,
such that the resulting constant $c \equiv \tfrac{1}{d}
\trace{}_{\sigma_n}\bigl( \hat{C} \bigr)$ is the plain average of
the operator $\hat{C}$ in the local basis $\vert \sigma_n \rangle$
that it acts upon. To be specific, this derives from the trace over
the environmental states $\vert e\rangle_n$ in \Eq{eq:FDM-wn}.
\EQ{eq:fdm-xvals} finally follows from \Eq{eq:FDM:xvals-2}, in that
for $n<n_0$, by construction, due to the absence of truncation the
second and third term in \Eq{eq:FDM:xvals-2} are identically zero.

\subsection{Time-dependent NRG}

Starting from the thermal equilibrium of some initial (\I)
Hamiltonian $\hat{H}^\I$, at time $t=0$ a quench at the location of
the quantum impurity occurs, with the effect that for $t>0$ the
time-evolution is governed by a different final (\F) Hamiltonian
$\hat{H}^\F$. While initially introduced within the single-shell
framework for finite temperature,\cite{Anders05} the same analysis
can also be straightforwardly generalized to the multi-shell
approach of \fdmNRG. Thus the description here will focus on the
\fdm approach.

Given a quantum quench, the typical time-dependent expectation value
of interest is
\begin{eqnarray}
   C(t) &\equiv& \bigl\langle \hat{C}(t) \bigr\rangle_T 
   \equiv
   \trace \bigl[ \rho^\I(T) \cdot e^{iH^\F t} \hat{C} e^{-iH^\F t}
   \bigr]
\text{, } 
\label{eq:tdm:C(t)}
\end{eqnarray}
with $\hat{C}$ some observable. While the physically relevant time
domain concerns the dynamics after the quench, \ie $t>0$, one is
nevertheless free to extend the definition of \Eq{eq:tdm:C(t)} also
to negative times. The advantage of doing so is, that the Fourier
transform into frequency space of the $C(t)$ in \Eq{eq:tdm:C(t)}
defined for arbitrary times becomes purely real, as will be shown
shortly. With this the actual time-dependent calculation can be
performed \emph{in frequency space first} in a simple and for the
NRG natural way,
\begin{eqnarray}
   C(\omega) &=& \int \tfrac{dt}{2\pi}  e^{i\omega t} \trace
   \bigl(
      \hat{\rho}^\I(T) \cdot
      e^{i\hat{H}^\F t} \hat{C} e^{-i\hat{H}^\F t}
   \bigr)
\text{ ,}\label{eq:tdm:C(w)}
\end{eqnarray}
A Fourier transform back into the time-domain at the end of the
calculation, finally, provides the desired time-dependent
expectation value $C(t) = \int C(\omega)e^{-i\omega t}\,d\omega$ for
$t\ge 0$. In order to obtain smooth data closer to the thermodynamic
limit, a weak log-Gaussian broadening in frequency space quickly
eliminates artificial oscillations in the time-domain which derive
from the logarithmic discretization. Note that for the sole purpose
of damping these artificial oscillations, typically a
\emph{significantly} smaller log-Gauss broadening parameter
$\alpha\lesssim 0.1$ suffices as compared to what is typically used
to obtain fully smoothened correlation functions in the frequency
domain, \eg $\alpha \gtrsim 0.5$ for $\Lambda=2$ (\eg see EPAPS of
Ref.~\onlinecite{Wb07}).

\begin{figure}[tb!]
\begin{center}
\includegraphics[width=\linewidth]{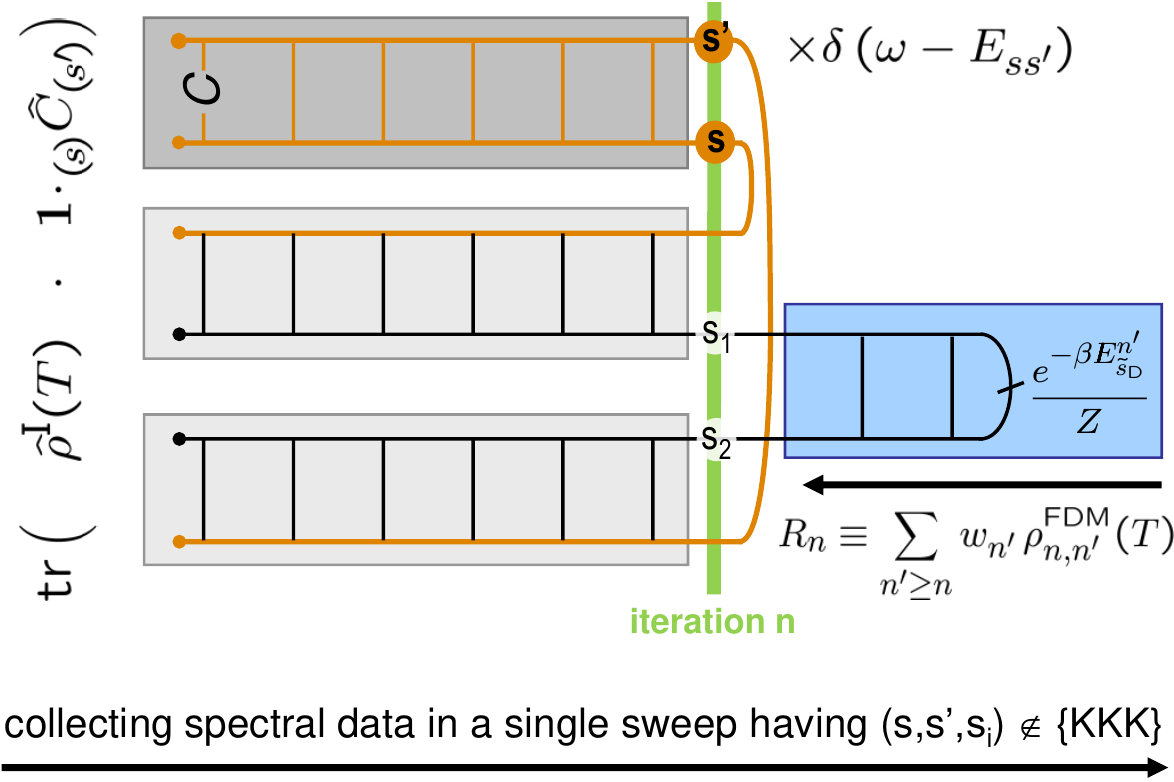}
\end{center}
\caption{ (Color online)
  MPS diagram for the calculation of quantum quenches using \tdmNRG
  [\cf \Eq{eq:tdm-spec}]. The calculation is performed in frequency
  space as depicted, which at the very end of the calculation is
  Fourier transformed back into the time domain to obtain the desired
  time-dependent expectation value $\hat{C}(t)$ [\cf
  \Eq{eq:tdm:C(w)}{}]. The calculation requires a complete eigenbasis
  for initial (black horizontal lines) and final Hamiltonian [orange
  (gray) horizontal lines], respectively, which are computed in two
  preceding NRG runs. Their respective shell-dependent overlap
  matrices $S_n$ (two light gray boxes at lower left) are calculated
  in parallel to the calculation of the matrix elements $C$ (dark gray
  box at the top). The projections $R_n^\I$ of the \fdm (box at the
  lower right) are evaluated with respect to the initial Hamiltonian,
  but have exactly the same structure otherwise as already discussed
  with \Fig{fig:fdmnrg} [see also \Eq{eq:fdm-Rn}]. The spectral data,
  finally, is collected in a full forward sweep, as indicated by the
  arrow at the bottom. To be specific, the summation is over all
  Wilson shells $n$ and for a given iteration $n$, over all states
  $(s,s',s_i) \notin \{\K\K\K\}$ with $s_i \in\{s_1,s_2\}$.
}\label{fig:tdmnrg}%
\end{figure}

\subsubsection{Lehmann representation}

For the Lehmann representation of \Eq{eq:tdm:C(w)}, in principle,
three complete basis sets are required: one completed basis set $i$
derived from an NRG run in $\hat{H}^\I$ to construct $\rho^\I(T)$,
and two complete basis sets $f$ and $f'$ from an NRG run in
$\hat{H}^\F$ to be inserted right before and after the $\hat{C}$
operator, respectively, to describe the dynamical behavior. Clearly,
two NRG runs in $\hat{H}^\I$ and $\hat{H}^\F$ are required to
describe the quantum quench. \cite{Oliveira81,Oliveira85,Helmes05}
With this, the spectral data in \Eq{eq:tdm:C(w)} becomes
\begin{eqnarray}
   C(\omega) &=& \sum_{i,f,f'}
   \underset{ \equiv S_{if'}^\ast }{\underbrace{
     \langle f'\vert i\rangle }}
   \,\rho_c^\I(T) \,
   \underset{ \equiv S_{if} }{\underbrace{ \langle i\vert f\rangle }}
   \,C_{ff'} \, \delta\left(\omega - E_{ff'}^\F\right)
\text{,}\label{eq:tdm:Lehmann}
\end{eqnarray}
which generates the overlap matrix $S$. Now using the complete NRG
eigenbasis sets together with the FDM, again similar to the \fdmNRG
in \Eq{eq:fdm-spec}, this introduces another sum over Wilson shells.
Therefore the fourier-transformed time-dependent NRG (\tdmNRG)
becomes,
\begin{eqnarray}
   C(\omega) &=& {\sum_{n,ss'}}'
        \bigl[ S^\dagger_n \, R_n^{\I,\tilde{\X}} S_n \bigr] _{s's}
        (C_n)_{ss'}
      \,\delta\left(\omega - E_{ss'}^n \right)
\text{,}\label{eq:tdm-spec}
\end{eqnarray}
where $(s,s') \in \{\X,\X'\}$. In addition, $\tilde{\X}
\in\{\K,\D\}$ describes the sector of the reduced density matrix
$R_n^{\I}$ from the initial system. To be specific, the notation for
the first term in \Eq{eq:tdm-spec} implies the matrix product,
\begin{eqnarray}
   \bigl[S^\dagger_n \, R_n^{\I,\tilde{\X}} S_n \bigr]^{\X'\X} \equiv
   \bigl(S_n^{\tilde{\X}\X'}\bigr)^\dagger \cdot R_n^{\I,\tilde{\X}}
   \cdot S_n ^{\tilde{\X}\X}
\text{.}\label{eq:tdm-spec:SRS}
\end{eqnarray}
For example, the left daggered overlap matrix $S_n$ selects the
overlap of the sectors $\{\tilde{\X},\X' \}$ between initial and
final eigenbasis, respectively. The prime in \Eq{eq:tdm-spec}
indicates that the sum includes all combinations of sectors
$\X\X'\tilde{\X} \neq KKK$, \ie a total of seven contributions. The
latter derives from the reduction of the independent three-fold sum
over Wilson shells [\cf \Eq{eq:tdm:Lehmann}] into a single sum over
Wilson shells $n$ as discussed with \Eq{NRG:redSumMto1}. It is
emphasized here, that the reduction of multiple sums in Wilson
shells as in \Eqs{NRG:redSum2to1}{NRG:redSumMto1} is \emph{not}
constrained to having the complete basis sets being identical to
each other. It is easy to see that it equally applies to the current
context of different basis sets from initial and final Hamiltonian.

The MPS diagram corresponding to \Eq{eq:tdm-spec} is shown in
\Fig{fig:tdmnrg}. It is similar to \Fig{fig:fdmnrg}, yet with
several essential differences: the block describing the matrix
elements of the original operator $\hat{B}$ has now become the block
containing $\hat{C}$. The original operator $\hat{C}^\dagger$ is
absent, \ie has become the identity. Yet since its ``matrix
elements'' are calculated with respect to two different basis sets
(initial and final Hamiltonian), an overlap matrix remains (lowest
block in \Fig{fig:tdmnrg}). In the context of the correlation
functions in \Fig{fig:fdmnrg}, the bra-ket states for the inserted
complete basis set in the index $s$ could be reduced to the single
bra-index $s$, such that it affected a single horizontal line only.
Here, however, two different complete basis sets hit upon each
other, which inserts another overlap matrix (second block from the
top in \Fig{fig:tdmnrg}, which corresponds to the Hermitian
conjugate of the lowest block). The reduced density matrices
$R_n^{\I,\X}$, finally, are built from the initial Hamiltonian, yet
are completely identical in structure otherwise to the ones already
introduced in \Eq{eq:fdm-Rn}.

The basis of the initial Hamiltonian enters through the two legs
(horizontal black lines)  in \Fig{fig:tdmnrg} which connect to the
density matrix $R_n$. All other legs refer to the NRG basis
generated by the final Hamiltonian [horizontal orange (gray) lines].
Finally, note that the plain contraction $S^\dagger R S$ of the
lower three tensors \wrt the indices $s_1$ and $s_2$ can simply be
evaluated through efficient matrix multiplication as in
\Eq{eq:tdm-spec:SRS}, while nevertheless respecting the selection
rules on the state space sectors $\{\X,\X',\tilde{\X}\} \neq
\{\K,\K,\K\}$.

\subsection{Fermi-Golden-Rule calculations}

The NRG is designed for quantum impurity models. As such, it is also
perfectly suited to deal with local quantum events such as
absorption or emission of a generalized local  impurity in contact
with non-interacting reservoirs. \cite{Oliveira81,Oliveira85,
Helmes05, Tureci11,Latta11,Wb11_aoc} If the rate of absorption is
weak, such that the system has sufficient time to equilibrate on
average, then the resulting absorption spectra are described by
Fermi's-Golden rule (fgr), \cite{Sakurai94}
\begin{eqnarray}
   A(\omega) &=& 2\pi \sum_{i,f} \rho_i^\I(T)
   \cdot \vert \langle f\vert \hat{C}^\dagger \vert i\rangle \vert^2
   \cdot \delta(\omega - E_{if})
\text{,}\label{eq:fgr-def}
\end{eqnarray}
where $i$ and $f$ describe complete basis sets for initial and final
system, respectively. The system starts in the thermal equilibrium
of the initial system. The operator $\hat{C}^\dagger$ describes the
absorption event at the impurity system, \ie corresponds to the term
in the Hamiltonian that couples to the light field. The transition
amplitudes between initial and final Hamiltonian are fully described
by the matrix elements $C_{if} \equiv \langle i\vert \hat{C} \vert
f\rangle$. Given that the energy difference $E_{if} \equiv E_f^{\F}
- E_i^{\I}$ in \Eq{eq:fgr-def} needs to be calculated between states
of initial and final system, absorption or emission spectra usually
show threshold behavior in the frequency $\omega$. The threshold
frequency is given by the difference in the ground state energies of
initial and final Hamiltonian, $\omega_\mathrm{thr} \equiv \Delta
E_g \equiv E_g^\F - E_g^\I$, which eventually is blurred by
temperature.

The difference between absorption and emission spectra is the
reversed role of initial and final system, while also having
$\hat{C}^\dagger \to \hat{C}$. That is, from the perspective of the
absorption process, the emission process starts in the thermal
equilibrium of the final Hamiltonian, with subsequent transition
matrix elements to the initial system. This also implies that
emission spectra have their contributions at negative frequencies,
\ie frequencies smaller than the threshold frequency
$\omega_\mathrm{thr}$ indicating the emission of a photon. Other
than that, the calculation of an emission spectrum is completely
analogous to the calculation of an absorption spectrum. With this in
mind, the following discussion will be therefore constrained to
absorption spectra only.

\begin{figure}[tb!]
\begin{center}
\includegraphics[width=\linewidth]{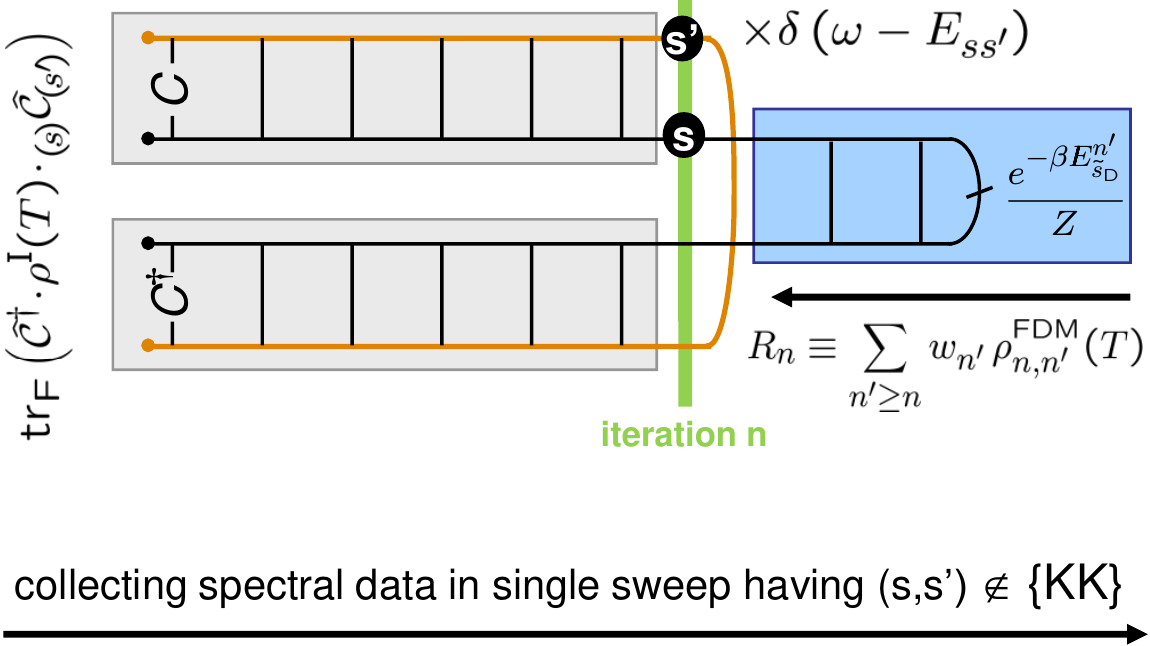}
\end{center}
\caption{(Color online)
  MPS diagram for the calculation of absorption spectra using
  Fermi's-Golden-rule (\fgrNRG) mediated by the operator
  $\hat{C}^\dagger$ [\cf \Eq{eq:fgr-spec}]. The two center legs
  (horizontal black lines) refer to the state space of the initial
  Hamiltonian, while the outer legs [horizontal orange (gray) lines]
  refer to the state space of the final Hamiltonian. Therefore the
  matrix elements of $\hat{C}^\dagger$ are \emph{mixed} matrix
  elements between eigenstates of initial and final Hamiltonian.
}\label{fig:fgrnrg}%
\end{figure}

While an absorption spectrum is already defined in frequency domain,
it nevertheless can be translated into the time domain through
Fourier transform,
\begin{eqnarray}
   A(t) &\equiv&
   \int \frac{d\omega}{2\pi} e^{-i \omega t}A(\omega) 
 \notag \\
 &=& \sum_{i,f} \rho_i^\I(T) \cdot e^{i E_{i} t}
   \langle i\vert \hat{C} \vert f\rangle e^{ -i E_{f} t} \cdot
   \langle f\vert \hat{C}^\dagger \vert i\rangle \notag\\
&=& \bigl\langle
   \underset{ \equiv \hat{C}(t) }
     {\underbrace{ e^{i \hat{H}^\I t} \hat{C} e^{-i \hat{H}^\F t}
      \raisebox{-0.05in}{\mbox{}} }}
   \cdot \hat{C}^\dagger \bigr\rangle _{T}^\I
\text{.}\label{eq:fgr-t}
\end{eqnarray}
Thus absorption spectra can also be interpreted similar to
correlation functions and quantum quenches: at time $t=0$ an
absorption event occurs (application of the operator
$\hat{C}^\dagger$, which for example rises an electron from a low
lying level into some higher level that participates in the
dynamics). This alters the Hamiltonian, such that the subsequent
time evolution is governed by the final Hamiltonian. At some time
$t>0$ then, the absorption event relaxes back to the original
configuration (application of $\hat{C}$). Therefore $A(t)$
essentially describes the overlap amplitude of the resulting state
with the original state with no absorption within the thermal
equilibrium of the initial system. While the ``mixed'' time
evolution of $\hat{C}(t)$ in \Eq{eq:fgr-t} may appear somewhat
artificial at first glance, it can be easily rewritten in terms of a
single time-independent Hamiltonian: by explicitly including a
further static degree (\eg a low lying hole from which the electron
was lifted through the absorption event, or the photon itself), this
switches $\hat{H}^\I$ to $\hat{H}^\F$, \ie between two dynamically
disconnected sectors in Hilbert space of the same Hamiltonian
(compare discussion of type-1 and type-2 quenches in
Ref.~\onlinecite{Muender12}).

Within the FDM formalism, the calculation of Fermi-Golden rule
calculations as defined in \Eq{eq:fgr-def} becomes (\fgrNRG),
\begin{eqnarray}
   A(\omega) &=& {\sum_{n,ss'}}'
        \bigl[ C_n^\dagger \, R_n^{\I} \bigr] _{s's}
        (C_n)_{ss'}
      \,\delta\left(\omega - E_{ss'}^n \right)
\text{,}\label{eq:fgr-spec}
\end{eqnarray}
where $(s,s')\in\{ \X^{\I}, \X^{\F}\}  \notin \{\K\K\}$. Therefore
$(C_n)_{ss'} \equiv {}^\I_n\langle s\vert \hat{C} \vert
s'\rangle^\F_n$ represents mixed matrix elements between states from
initial and final Hamiltonian, respectively, which nevertheless can
also be easily calculated using the basic contractions discussed
with \Fig{fig:mps:basics}.

The MPS diagram to be evaluated for \Eq{eq:fgr-spec} is shown in
\Fig{fig:fgrnrg}. Its structure is completely analogous to the
calculation of generic correlation functions in \Fig{fig:fdmnrg},
except that similar to the quantum quench earlier, here again the
basis sets from two different Hamiltonians come into play.
\cite{Oliveira81,Oliveira85,Helmes05} In contrast to the quantum
quench situation in \Fig{fig:tdmnrg}, however, no explicit overlap
matrices are required. Instead, all matrix elements of the local
operator $\hat{C}^\dagger$ themselves are mixed matrix elements
between initial and final system.
The reduced density matrices $R_n^{\I}$ are constructed \wrt the
initial Hamiltonian, but again exactly correspond to the ones
already introduced in \Eq{eq:fdm-Rn} otherwise.

\subsubsection{Technical remarks}

Absorption or emission spectra in the presence of Anderson
orthogonality or strongly correlated low-energy physics typically
exhibit sharply peaked features close to the threshold frequency
with clear physical interpretation. While in principle, a single
Hamiltonian with dynamically disconnected Hilbert space sectors may
have been used, this is ill-suited for an NRG simulation. Using a
single NRG run, this can only resolve the low-energy of the full
Hamiltonian, \ie of the initial system as it is assumed to lie lower
in energy. Consequently, the sharp features at the threshold
frequency will have to be smoothened by an energy window comparable
to $\omega_\mathrm{thr} = \Delta E_g$ in order to suppress
discretization artifacts. This problem is fully circumvented only by
using two separate NRG runs, one for the initial and one for the
final Hamiltonian. \cite{Oliveira81,Oliveira85,Helmes05}
With the NRG spectra typically collected in logarithmically spaced
bins, having two NRG runs then, it is important that the data is
collected in terms of the frequencies $\nu \equiv \omega -
\omega_\mathrm{thr}$ taken \emph{relative to the threshold
frequency} $\omega_\mathrm{thr}$ as defined earlier.

\begin{figure}[tb!]
\begin{center}
\includegraphics[width=0.9\linewidth]{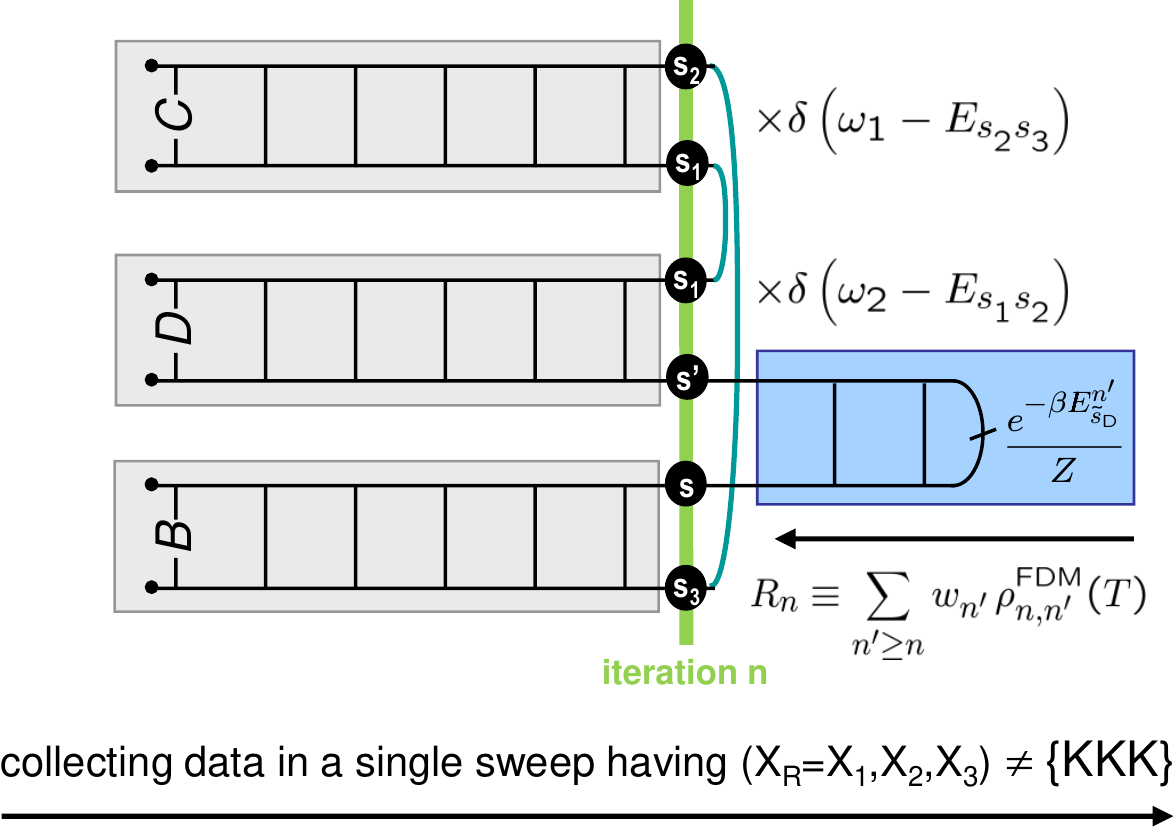}
\end{center}
\caption{
  MPS diagram for the evaluation of a three-point correlation
  functions as in \Eq{eq:fdmnrg:2}, which thus generalizes
  \fdmNRG (see also \Fig{fig:fdmnrg}).
}\label{fig:fdmnrg2}%
\end{figure}

\subsection{Higher-order correlation functions}

Consider the spectral function of a three-point correlation
function, which in the time domain is given by
\begin{align}
    A_{BCD}(t_1,t_2)
 &\equiv \langle \hat{D}(t_2) \hat{C}(t_1) \hat{B}\rangle_T
    \label{eq:fdmnrg:2t} \\
 &\equiv \trace\Bigl(
       \hat{\rho}(T)\cdot
       e^{i\hat{H}t_2} \hat{D} e^{i\hat{H}(t_1-t_2)} \hat{C}
       e^{-i\hat{H}t_1} \hat{B}
    \Bigr)
\notag
\end{align}
Given a time-invariant Hamiltonian, the correlator of three
operators $\hat{B}$, $\hat{C}$, and $\hat{D}$ acting at three
different times results in the dependence on effectively two times
$t_1$ and $t_2$, since $t_0$ as in $\hat{B}(t_0)$ can simply be
chosen arbitrary, \ie $t_0=0$ for simplicity.

Using the NRG eigenbasis sets together with the FDM, the Lehman
representation of \Eq{eq:fdmnrg:2t} requires four independent sums
over complete eigenbasis sets, one from the \fdm ($\X_\rho$), and
three by inserting an identity with every exponentiated Hamiltonian
($\X_1,\X_2,\X_3$ from left to right), respectively. Again, with the
reduced density matrix $\rho$ being a scalar operator, one has
$\X_\rho=\X_1$. Using \Eq{NRG:redSumMto1} then, in frequency space
\Eq{eq:fdmnrg:2t} becomes
\begin{align}
    A_{BCD}(\omega_1,\omega_2)
  =& \sum_{n}{\sum_{s_1 s_2 s_3}}^{\!\!\!\prime}
   \ \bigl[B_n R_n \bigr]_{s_3,s_1} (D_n)_{s_1 s_2} (C_n)_{s_2 s_3}
   \notag\\
  &\times
   \,\delta(\omega_2-E_{s_1,s_2}^n)
   \,\delta(\omega_1-E_{s_2,s_3}^n)
\text{,}\label{eq:fdmnrg:2}
\end{align}
with $(s_1,s_2,s_3) \in \{ \X_1,\X_2,\X_3 \}\neq \{\K\K\K\}$, as
indicated by the prime next to the sum, having $\bigl[B_n R_n
\bigr]^{\X_3,\X_1} \equiv B_n ^{\X_3,\X_1} R_n^{\X_1}$. The MPS
diagram corresponding to the spectral representation in
\Eq{eq:fdmnrg:2} is shown in \Fig{fig:fdmnrg2}.

The more challenging part with \Eq{eq:fdmnrg:2} is the dependence on
two frequencies. While the corresponding full collection of data
into bins $(\omega_1,\omega_2)$ can become expensive, however,
certain fixed frequency points together with different kernels
corresponding to a different analytic structure of the higher-order
correlation function [which then replace the $\delta$-functions in
\Eq{eq:fdmnrg:2}], appear feasible with reasonable effort. Moreover,
within the NRG context, by construction, one has comparable energy
resolution for $\omega_1$ and $\omega_2$ at a given energy shell.
Hence it remains to be seen in what respect vastly different energy
scales of $\omega_1$ as compared to $\omega_2$, if required, are
affected by the ansatz of energy scale separation within the NRG.

\section{Summary and conclusions \label{sec:summary}}

The framework of tensor network has been applied to the NRG. This
makes full use of the complete basis sets as introduced by
\mycite{Anders05} which, within the approximation of energy scale
separation, also represent many-body eigenstates of the full
Hamiltonian. Together with the full density matrix (\FDM) approach,
complete basis sets allow for simple transparent algorithms, as
demonstrated for correlation function (\fdmNRG), time-dependent
quenches (\tdmNRG), as well as Fermi-Golden-rule (\fgrNRG)
calculations. The underlying principle is based on the plain Lehmann
representation of the relevant dynamical expressions, which within
the NRG, can be evaluated in a text-book like clean and transparent
fashion.

The framework of complete basis sets clearly allows for
straightforward further generalizations. For example, one can
envisage multiple consecutive time-steps which thus generalizes
\tdmNRG with the possibility to implement periodic switching.
\cite{Eidelstein12} While initially, the system starts in thermal
equilibrium of a given Hamiltonian, after each quench the
description of the system must be projected onto the complete basis
set of the following Hamiltonian in terms of the reduced density
matrix. For all these calculations, however, one must keep in mind
that Wilson chains are not thermal reservoirs. \cite{Rosch11} Within
the \tdmNRG, for example, this can manifest itself as \emph{finite
size effect}, in that already for a single quench in the absence of
an external magnetic field, an initial excess spin at the impurity
\emph{cannot} be fully dissipated into the bath even in the limit of
time $t\to\infty$, leading to (small) residual magnetization at the
impurity. 
In cases where these discretization effects become a strong limiting
factor, hybrid NRG approaches have been devised with the idea to
extend the bath to a more refined or uniform spectrum. However,
since this typically compromises energy scale separation along the
full Wilson chain, other methods such as the DMRG need to be
incorporated. \cite{Wb09,Guettge12}

\begin{acknowledgements}

I would like to thank Theo Costi, Markus Hanl, and Jan von Delft for
their critical reading of the manuscript. This work has received
support from the German science foundation (TR-12, SFB631, NIM, and
WE4819/1-1).

\end{acknowledgements}

\appendix

\section{Fermionic signs}

The NRG is typically applied to fermionic systems (for extensions to
bosonic systems see, for example, Refs.~[\onlinecite{Bulla03,
Bulla08, Guo12}]). Through its iterative prescription, the resulting
MPS has a specific natural fermionic order in Fock space,
\begin{align}
  \vert s\rangle_n &= \hspace{-0.1in}
  \sum_{\sigma_d,\sigma_0,\ldots,\sigma_n} \bigl(
      A^{[\sigma_d]} A^{[\sigma_0]} \cdot \ldots \cdot
      A^{[\sigma_n]}
  \bigr)_s
    \underset{
      \equiv \vert \sigma_n, \ldots, \sigma_0, \sigma_d \rangle
    }{\underbrace{
      \vert \sigma_n\rangle \ldots
      \vert \sigma_0\rangle
      \vert \sigma_d\rangle
    }}
\text{,}\label{eq:mps:sn}
\end{align}
where $\vert \sigma_d\rangle$ stands for the local state space of
the impurity. Site $n'>n$ is added after site $n$, hence the state
space $\vert \sigma_{n'}\rangle$ naturally appears to the left
$\vert \sigma_{n}\rangle$ with second quantization in mind. The
environmental states $\vert e\rangle_n$ \wrt to iteration $n$ which
refers to the sites $n'>n$ is irrelevant for the following
discussion, and hence will be skipped.

Let $\hat{c}^\dagger$ be a fermionic operator that acts on the
impurity. Here $\hat{c}^\dagger$ is assumed an arbitrary operator
that nevertheless creates or destroys an odd number of fermionic
particles such that fermionic signs apply. A very frequent task then
is to represent this operator in the effective many-body-basis at
iteration $n$, \ie to calculate the matrix elements
$(C_n^\dagger)_{ss'} \equiv {}_{n\!}\langle s \vert \hat{c}^\dagger
\vert s' \rangle_n$ [\cf \Fig{fig:mps:basics}]. This involves the
basic matrix-element \wrt to local state spaces,
\begin{align}
 & \langle \sigma_n,\ldots,\sigma_0,\sigma_d\vert \hat{c}^\dagger
   \vert \sigma'_n,\ldots,\sigma'_0,\sigma'_d\rangle =
  \notag\\
 & \qquad = \Bigl[\prod_{i=n,\ldots,0}
    \underset{ \equiv (\hat{z}_i)_{\sigma_i,\sigma'_i} }{\underbrace{
        \Bigl( \delta_{\sigma_i,\sigma'_i}  (-1)^{n_{\sigma'_i}} \Bigr)
    }}
    \Bigr] \cdot
    \langle \sigma_d\vert  \hat{c}^\dagger \vert \sigma'_d\rangle
\text{,}\label{eq:sn}
\end{align}
with $\hat{z} \equiv (-1)^{\hat{n}} = \exp(i\pi \hat{n})$, akin to
the Pauli z-matrix. That is, by pulling the operator
$\hat{c}^\dagger$, acting on the impurity, to the right
past the second quantization operators that create the states
$\sigma_{n_i}$, fermionic signs apply, resulting in a
\emph{Jordan-Wigner string}
\begin{equation}
  \hat{Z} \equiv \bigotimes\limits_{i=0,\ldots,n} \hat{z}_i
\text{,}\label{eq:mps:zstring}
\end{equation}
to be called \emph{z-string} in short. Note that through the
Jordan-Wigner transformation, which maps fermions onto spins and
vice versa, exactly the same string operator as in
\Eq{eq:mps:zstring} emerges. For a one-dimensional system with
nearest neighbor hopping, the Jordan-Wigner transformation to spins
allows to eliminate on the operator level of the Hamiltonian further
complications with fermionic signs. This is fully equivalent, of
course, to the explicit treatment of the Jordan-Wigner string in a
numerical setting that keeps a fermionic basis.
The operators $\hat{z}_i$ in \Eq{eq:sn} take care of the book
keeping of fermionic signs, by inserting $-1$ ($+1$) for all states
$\sigma_i$ at site $i$ with odd (even) number of particles
$n_{\sigma_i}$. The operators $\hat{z}_i$ are diagonal and hence
commute with each other. In the case of additional explicit
spin-degrees of freedom, such as the localized spin in the Kondo
model, its z-operator is proportional to the identity matrix and
hence can be safely ignored.

In the following, three alternative viewpoints are discussed for
dealing with fermionic signs in the MPS setup of the NRG. To be
specific, the following discussion assumes $\hat{c}^\dagger =
\hat{d}^\dagger$ which creates a particle at the impurity's
$d$-level. As such, it generates a Jordan Wigner string for all
sites added subsequently to the MPS, \ie sites $i=0,\ldots,n$ [\cf
\Eq{eq:mps:zstring}].

\begin{figure}[tb!]
\begin{center}
\includegraphics[width=\linewidth]{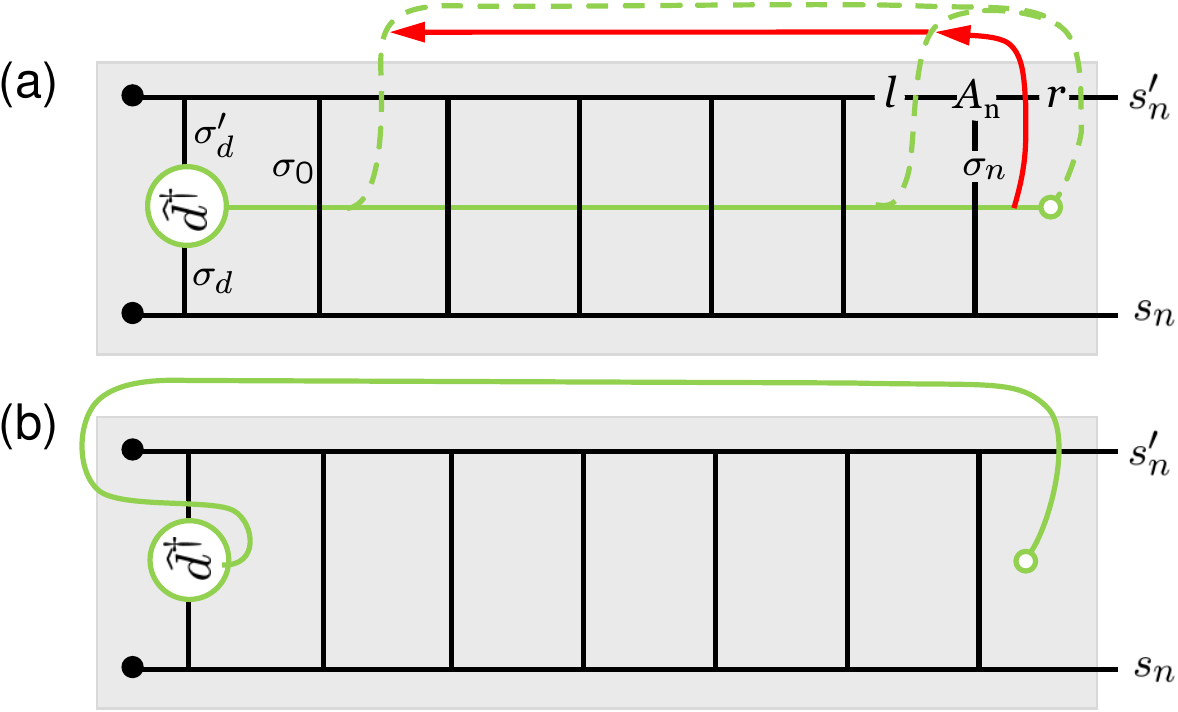}
\end{center}
\caption{ (Color online)
  MPS diagrams and fermionic signs. Consider the matrix elements of a
  local operator $\hat{d}^\dagger$ which creates a particle at the
  impurity, \ie the first local state space of the MPS in the
  effective MPS space $\vert s\rangle_n$. A z-string
  (Jordan-Wigner string) $\hat{Z}= \bigotimes_i \hat{z}_i$ arises
  [green (gray) horizontal line in the middle]. The endpoints (open
  circles) indicate the range of the z-string, \ie starting from and
  including site $0$ to site $n$. For every crossing of the z-string
  with a black line, which represent state spaces, fermionic signs
  apply. Panel (a) shows that a z-string can be rerouted (light dashed
  lines, pushed in the direction of the red arrow). The resulting
  configuration in panel (b) shows that by rerouting the z-string
  significantly fewer crossings with black lines can be achieved. In
  particular, the z-strings which applied to all sites to the right of
  $\hat{d}^\dagger$ (panel a), can be significantly reduced to local
  fermionic signs at the impurity and another fermionic sign with the
  state space $s_n$.
}\label{fig:zstrings1}%
\end{figure}

\subsubsection*{Viewpoint 1: rerouting of z-string in tensor network}

\FIG{fig:zstrings1} depicts an MPS diagram for the typical
evaluation of matrix elements with relevant fermionic signs. The
\Atensors that derive from a preceeding iterative state space
generation of the NRG are depicted by the ternary nodes (\cf
\Fig{fig:mps:basics}). By keeping track of the total number $n$ of
particles for all indices then, for some specific index $a$ the
fermionic sign is given by $(-1)^{n_a}$.

The z-string that is required for the evaluation of the matrix
elements of $d^\dagger$, stretches across all local state spaces
$\sigma_{i}$ with $0\leq i\leq n$. This is depicted by the light
green (gray) line in \Fig{fig:zstrings1} (note that this is not the
extra index that takes care of non-abelian symmetries as in
\Fig{fig:mps:basics2},  even though graphically its role is not that
dissimilar). Here the interpretation is such, that a \emph{crossing}
of the z-string with a state space inserts fermionic signs for this
state space. \cite{Corboz10,Kraus10,Barthel09} Consider then, for
example, the upper right \Atensor, $A_n$, in \Fig{fig:zstrings1}.
For simplicity, its three legs are labeled $l \equiv s_{n-1}$ (state
space from previous iteration), $\sigma_{(n)}$ (new local state
space), and $r\equiv s_n$ (combined state space) for left, local,
and right, respectively. By tracking the total particle number for
all states, given the left-to-right orthonormalization (see arrows
in \Fig{fig:mps:basics}), by construction it must hold
$n_l+n_\sigma=n_r$. The index $\sigma$ is \emph{crossed} by the
z-string, hence fermionic signs apply at the location of the
crossing,
\begin{eqnarray}
   z_\sigma \equiv (-1)^{n_\sigma} = (-1)^{n_r}
   \underset{ =(-1)^{+n_l} }{\underbrace{ (-1)^{-n_l} }}
   \equiv z_l z_r
\text{.}\label{eq:z-string:nrule}
\end{eqnarray}
Therefore, instead of applying fermionic signs with index $\sigma$,
it is equally correct to apply fermionic signs with the indices $l$
and $r$. This allows to \emph{reroute} the z-string
\cite{Corboz10,Kraus10,Barthel09} as indicated in
\Fig{fig:zstrings1} (dashed line to the upper right, with the shift
in the z-string indicated by short red arrow). Note that for this
rerouting to work, the actual left-to-right orthonormalization is
not strictly required, and could be relaxed, in general, to the more
general condition $n_l \pm n_r \pm n_\sigma = \mathrm{even}$. In
particular, this includes $n_l \pm n_r \pm n_\sigma = 0$, which
suggests that \emph{any} direction of orthonormalization is
acceptable, together with a generic \emph{current site} that
combines all (effective) state spaces to an even number of
particles, \ie $n_l + n_r + n_\sigma = n_\mathrm{tot}=\mathrm{even}$
(for $n_\mathrm{tot}=\mathrm{odd}$, a global minus sign would apply
in the case of rerouting).

The basic rerouting step as indicated above can be repeated, such
that the z-string can be pulled to the top outside the MPS diagram
in \Figp{fig:zstrings1}{a}, with the final configuration shown in
\Figp{fig:zstrings1}{b}. The state to the very left (black dot) is
the vacuum states with no particles, hence the z-string can be fully
pulled outside at the left. As a result, instead of the original $n$
crossings with the state space $\sigma_n$, only two crossings of the
z-string with state spaces (black lines) remain: one crossing with
the local state space at the impurity itself, leading to
\begin{eqnarray}
   \hat{d}^\dagger \to \hat{d}^\dagger \hat{z}_d
   \equiv (\hat{z} \hat{d} )^\dagger
\text{,}\label{eq:z-string:drule}
\end{eqnarray}
which fully acts within the state space of the impurity, and another
crossing with the state space $\vert s'\rangle_n$ at iteration $n$.

\begin{figure}[tb!]
\begin{center}
\includegraphics[width=\linewidth]{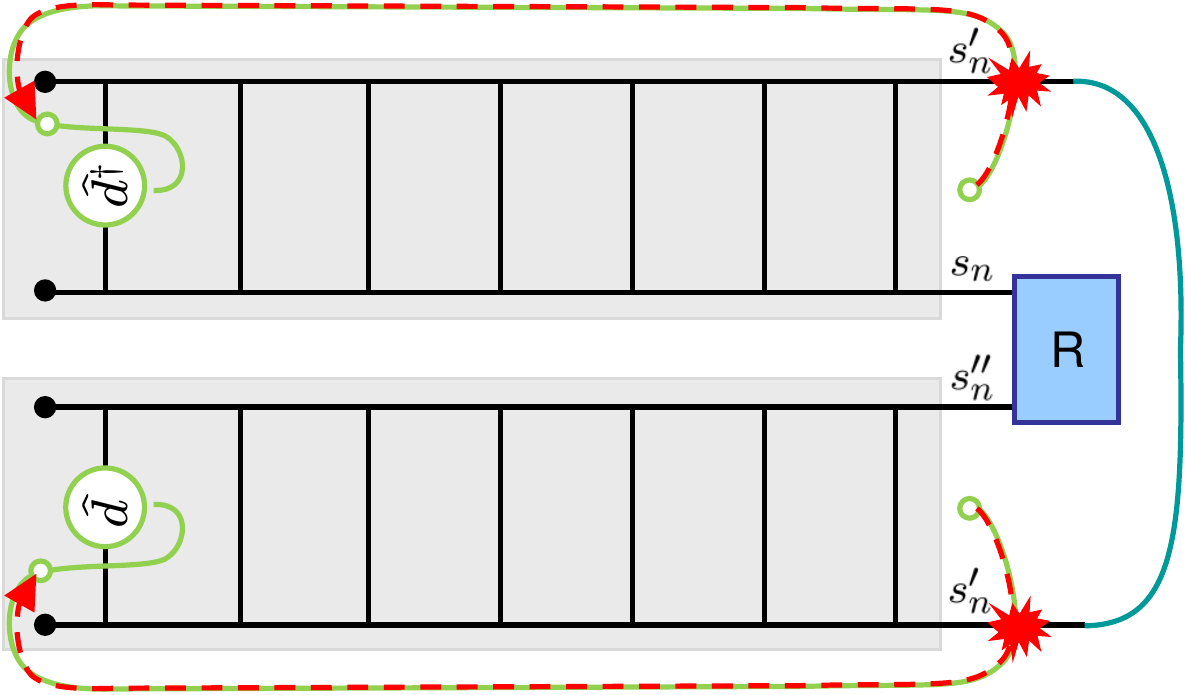}
\end{center}
\caption{(Color online)
  Example: fermionic signs in correlation functions. Two MPS diagrams
  as in \Fig{fig:zstrings2} for the matrix elements of $\hat{d}$ and
  $\hat{d}^\dagger$ are combined, as required, for example, for the
  calculation of correlation functions. The resulting product of
  matrix elements ${}_n\langle s'\vert \hat{d}\vert s''\rangle_n \cdot
  R_{s'',s}^{[n]}\cdot {}_n\langle s\vert \hat{d}^\dagger \vert
  s'\rangle_n$ leads to cancelation of the fermionic signs in the
  index $s'$ in the rerouted z-strings (light green lines), as
  indicated by the two splashes to the right. Hence the right
  end-point of the z-string can be fully retracted to the very left
  of the diagram, as indicated by the dashed red arrows. The partial
  contribution $R$ to the \fdm is a scalar operator, such that
  assuming charge conservation, the particle number of the states $s$
  and $s''$ also must be same. Hence the z-string in
  \Fig{fig:zstrings2} could have been equally well also rerouted
  downwards, instead. The respective fermionic signs with states $s$
  and $s''$ still would have canceled, while the order of application
  of the z-operator with the impurity would have changed.
}\label{fig:zstrings2}%
\end{figure}

In typical applications which include thermal expectation values or
correlation functions, however, an operator $\hat{d}^\dagger$ never
comes by itself, as its expectation value with respect to any state
with well-defined particle number would be zero. Therefore creation
and annihilation operators always appear in pairs. For the local
spectral function, for example, $\hat{d}^\dagger$ is paired with its
daggered version $\hat{d}$. In their overall combination, the
fermionic signs \wrt the index $s'$ appear \emph{twice} and hence
annihilate each other. This situation is sketched in
\Fig{fig:zstrings2}. The matrix element discussed previously with
\Fig{fig:zstrings1} is shown in the upper half of the figure. Given
the case of spectral functions (\cf \Fig{fig:fdmnrg}), its
counterpart is shown at the bottom. The reduced density matrix $R$
is a scalar operator, such that the particle number of the states
$s$ and $s''$ must match. Similarly, the outer two states are
connected through the overall trace (solid line to the very right),
hence refer to exactly the same state. Consequently, the same
fermionic sign factor applies twice with the rerouted z-strings,
which thus cancels, \ie $[ (-1)^{n_s} ]^2 = 1$ (indicated by the two
splashes with $s'$ at the right). Consequently, the right end-point
of the z-strings can be retracted along the rerouted z-string all to
the way to the left of the impurity (indicated by the red dashed
arrow).

Therefore given the \Atensors for the basis transformations from a
prior NRG run that only generates the basis, above line of argument
allows to ignore fermionic signs for most of the subsequent
calculation of thermodynamic quantities or spectral properties.
Specifically, in given example which applies to \fdmNRG, \tdmNRG, as
well as \fgrNRG, it is sufficient to calculate the spectral
functions for the operator $\hat{d}\to \hat{z}_d \hat{d}$ [\cf
\Eq{eq:z-string:drule}{}] and \emph{fully ignore} fermionic signs
for the rest of the chain. This is in contrast to the original setup
where the full z-string needs to be included.

\subsubsection*{Viewpoint 2: Operator representation}

An alternative way to demonstrate the effect of rerouting of the
z-string can be given by looking at the equivalent (numerical)
tensor-product representation of operators in the full many-body
Hilbert space without making reference to MPS notation. Given the
fermionic order of sites as in \Eq{eq:mps:sn}, a fermionic operator
$\hat{c}_k$ that destroys a particle at site $k$, has the
tensor-product form
\begin{eqnarray}
   \hat{C}_k \equiv
   \hat{1}_d\otimes \hat{1}_0\otimes \ldots \hat{1}_{k-1}\otimes
   \hat{c}_k\otimes \hat{z}_{k+1} \otimes\ldots\otimes \hat{z}_n
\text{,}\label{eq:fop-rep0}
\end{eqnarray}
where $\hat{1}_i$ is the identity matrix at site $i$, $\hat{c}_k$
the desired operator acting within the state space of site $k$, and
$\hat{z}_i \equiv (-1)^{\hat{n}_i}$ as in \Eq{eq:sn}. Now, applying
a z-operator to the states $s'$ at the last site $n$ is equivalent
to applying a z-operator to each individual site,
\begin{eqnarray}
   \hat{Z} \hat{C}_k
&\equiv& \Bigl(\bigotimes\limits_{i=d}^n \hat{z}_i\Bigr) \hat{c}_k \notag\\
&=& \hat{z}_d\otimes \hat{z}_0\otimes \ldots \hat{z}_{k-1}\otimes
   [\hat{z}\hat{c}]_k\otimes \hat{1}_{k+1}\ldots \hat{1}_n
\text{,}\label{eq:fop-rep}
\end{eqnarray}
since $(\hat{z}_i)^2=\hat{1}_i$. In the application to thermodynamic
quantities such as correlations functions, the operator $\hat{C}_k$
would again appear together with its daggered version
$\hat{C}_k^\dagger$, hence insertion of $\hat{Z}^2$ has no effect,
yet can be split in equal parts, \ie $\hat{C}_k^\dagger \hat{C}_k =
(\hat{Z} \hat{C}_k)^\dagger (\hat{Z} \hat{C}_k)$. Therefore,
$\hat{Z} \hat{C}_k$ can be equally well used instead of $\hat{C}_k$.
As a result, similar to \Fig{fig:zstrings2}, the z-strings have
again been fully flipped from the sites to the right of site $k$ to
the left of site $k$, with the additional transformation $\hat{c}_k
\to [\hat{z}\hat{c}]_k$. Note that, essentially, this equivalent to
fully reverting the fermionic order.

\subsubsection*{Viewpoint 3: Auxiliary fermionic level}

In the case of absorption spectra, the absorption of a photon
creates an electron-hole pair, $\hat{h}^\dagger\hat{d}^\dagger$,
where the hole $\hat{h}^\dagger$ can be simply treated as a
spectator in the dynamics. Nevertheless, by explicitly including the
hole in the correlation function, \ie by using the operator
$\hat{d}^\dagger \to \hat{h}^\dagger \hat{d}^\dagger$, this operator
itself now already forms a pair of fermions that preserves particle
number (assuming that $\hat{h}^\dagger$ creates a hole). Therefore,
by construction, $\hat{h}^\dagger\hat{d}^\dagger$ simply commutes
with all Wilson sites except for the impurity upon which it acts.

The same argument can be repeated for a standard spectral function,
by introducing an auxiliary fermionic level $\hat{h}$ that does not
participate in the dynamics, \ie does not appear in the Hamiltonian.
In general, \emph{prepending} the states in \Eq{eq:mps:sn} by the
states $\vert \sigma_h\rangle$ of the ``hole'', \ie
\begin{eqnarray}
    \vert \sigma_n, \ldots, \sigma_0, \sigma_d\rangle \to
    \vert \sigma_n, \ldots, \sigma_0, \sigma_d\rangle
    \vert \sigma_h\rangle
\text{,}\label{eq:mps:snh}
\end{eqnarray}
immediately results in the same consistent picture as already
encountered with \Fig{fig:zstrings2} or \Eq{eq:fop-rep}.

\bibliographystyle{unsrtnat} 
\bibliography{D:/TEX/Lib/mybib}

\end{document}